\documentclass[%
 reprint,
superscriptaddress,
 amsmath,amssymb,
 aps,
]{revtex4-2}

\usepackage{physics}
\usepackage{color}
\usepackage{multirow}
\usepackage{gensymb}
\usepackage{graphicx}
\usepackage{multirow}
\usepackage{dcolumn}
\usepackage{bm}
\usepackage[colorlinks,linkcolor=black,anchorcolor=black,citecolor=black]{hyperref}
\usepackage{float}

\begin{document}

\preprint{APS/123-QED}

\title{High-precision pulse calibration of tunable couplers for high-fidelity two-qubit gates\\ in superconducting quantum processors}

\author{Tian-Ming Li}
    \thanks{These authors contributed equally to this work.}
\author{Jia-Chi Zhang}
    \thanks{These authors contributed equally to this work.}
\author{Bing-Jie Chen}
    \thanks{These authors contributed equally to this work.}
    \affiliation{Institute of Physics, Chinese Academy of Sciences, Beijing 100180, China}
    \affiliation{School of Physics, University of Chinese Academy of Sciences Beijing 100049, China}
    
\author{Kaixuan Huang}
    \affiliation{Beijing Academy of Quantum Information Sciences, 100193 Beijing, China}
    
\author{Hao-Tian Liu}
    \affiliation{Institute of Physics, Chinese Academy of Sciences, Beijing 100180, China}
    \affiliation{School of Physics, University of Chinese Academy of Sciences Beijing 100049, China}
    
\author{Yong-Xi Xiao}
    \affiliation{Institute of Physics, Chinese Academy of Sciences, Beijing 100180, China}
    \affiliation{School of Physics, University of Chinese Academy of Sciences Beijing 100049, China}
\author{Cheng-Lin Deng}
    \affiliation{Institute of Physics, Chinese Academy of Sciences, Beijing 100180, China}
    \affiliation{School of Physics, University of Chinese Academy of Sciences Beijing 100049, China}
\author{Gui-Han Liang}
    \affiliation{Institute of Physics, Chinese Academy of Sciences, Beijing 100180, China}
    \affiliation{School of Physics, University of Chinese Academy of Sciences Beijing 100049, China}
\author{Chi-Tong Chen}
    \affiliation{Quantum Science Center of Guangdong-Hong Kong-Macao Greater Bay Area, 518045 Shenzhen, Guangdong, China}
\author{Yu Liu}
    \affiliation{Institute of Physics, Chinese Academy of Sciences, Beijing 100180, China}
    \affiliation{School of Physics, University of Chinese Academy of Sciences Beijing 100049, China}
    
\author{Hao Li}
    \affiliation{Beijing Academy of Quantum Information Sciences, 100193 Beijing, China}
\author{Zhen-Ting Bao}
    \affiliation{Institute of Physics, Chinese Academy of Sciences, Beijing 100180, China}
    \affiliation{School of Physics, University of Chinese Academy of Sciences Beijing 100049, China}
\author{Kui Zhao}
    \affiliation{Beijing Academy of Quantum Information Sciences, 100193 Beijing, China}
\author{Yueshan Xu}
    \affiliation{Beijing Academy of Quantum Information Sciences, 100193 Beijing, China}
\author{Li Li}
    \affiliation{Institute of Physics, Chinese Academy of Sciences, Beijing 100180, China}
    \affiliation{School of Physics, University of Chinese Academy of Sciences Beijing 100049, China}
\author{Yang He}
    \affiliation{Institute of Physics, Chinese Academy of Sciences, Beijing 100180, China}
    \affiliation{School of Physics, University of Chinese Academy of Sciences Beijing 100049, China}
\author{Zheng-He Liu}
    \affiliation{Institute of Physics, Chinese Academy of Sciences, Beijing 100180, China}
    \affiliation{School of Physics, University of Chinese Academy of Sciences Beijing 100049, China}
\author{Yi-Han Yu}
    \affiliation{Institute of Physics, Chinese Academy of Sciences, Beijing 100180, China}
    \affiliation{School of Physics, University of Chinese Academy of Sciences Beijing 100049, China}
\author{Si-Yun Zhou}
    \affiliation{Institute of Physics, Chinese Academy of Sciences, Beijing 100180, China}
    \affiliation{School of Physics, University of Chinese Academy of Sciences Beijing 100049, China}
\author{Yan-Jun Liu}
    \affiliation{Institute of Physics, Chinese Academy of Sciences, Beijing 100180, China}
    \affiliation{School of Physics, University of Chinese Academy of Sciences Beijing 100049, China}
\author{Xiaohui Song}
    \affiliation{Institute of Physics, Chinese Academy of Sciences, Beijing 100180, China}
    \affiliation{School of Physics, University of Chinese Academy of Sciences Beijing 100049, China}
    
\author{Dongning Zheng}
    \affiliation{Institute of Physics, Chinese Academy of Sciences, Beijing 100180, China}
    \affiliation{School of Physics, University of Chinese Academy of Sciences Beijing 100049, China}
    \affiliation{Beijing Academy of Quantum Information Sciences, 100193 Beijing, China}
    \affiliation{Hefei National Laboratory, 230088 Hefei, China}
    
\author{Zhongcheng Xiang}
    \thanks{zcxiang@iphy.ac.cn}
    \affiliation{Institute of Physics, Chinese Academy of Sciences, Beijing 100180, China}
    \affiliation{School of Physics, University of Chinese Academy of Sciences Beijing 100049, China}
    \affiliation{Beijing Academy of Quantum Information Sciences, 100193 Beijing, China}
    \affiliation{Hefei National Laboratory, 230088 Hefei, China}

\author{Yun-Hao Shi}
    \thanks{yhshi@iphy.ac.cn}
    \affiliation{Institute of Physics, Chinese Academy of Sciences, Beijing 100180, China}
    \affiliation{School of Physics, University of Chinese Academy of Sciences Beijing 100049, China}
    
\author{Kai Xu}
    \thanks{kaixu@iphy.ac.cn}
    \affiliation{Institute of Physics, Chinese Academy of Sciences, Beijing 100180, China}
    \affiliation{School of Physics, University of Chinese Academy of Sciences Beijing 100049, China}
    \affiliation{Beijing Academy of Quantum Information Sciences, 100193 Beijing, China}
    \affiliation{Hefei National Laboratory, 230088 Hefei, China}
    \affiliation{Songshan Lake Materials Laboratory, 523808 Dongguan, Guangdong, China}

\author{Heng Fan}
    \thanks{hfan@iphy.ac.cn}
    \affiliation{Institute of Physics, Chinese Academy of Sciences, Beijing 100180, China}
    \affiliation{School of Physics, University of Chinese Academy of Sciences Beijing 100049, China}
    \affiliation{Beijing Academy of Quantum Information Sciences, 100193 Beijing, China}
    \affiliation{Hefei National Laboratory, 230088 Hefei, China}
    \affiliation{Songshan Lake Materials Laboratory, 523808 Dongguan, Guangdong, China}


\begin{abstract}
\noindent

 For superconducting quantum processors, stable high-fidelity two-qubit operations depend on precise flux control of the tunable coupler. However, the pulse distortion poses a significant challenge to the control precision. Current calibration methods, which often rely on microwave crosstalk or additional readout resonators for coupler excitation and readout, tend to be cumbersome and inefficient, especially when couplers only have flux control. Here, we introduce and experimentally validate a novel pulse calibration scheme that exploits the strong coupling between qubits and couplers, eliminating the need for extra coupler readout and excitation. Our method directly measures the short-time and long-time step responses of the coupler flux pulse transient, enabling us to apply predistortion to subsequent signals using fast Fourier transformation and deconvolution. This approach not only simplifies the calibration process but also significantly improves the precision and stability of the flux control. We demonstrate the efficacy of our method through the implementation of diabatic CZ and iSWAP gates with fidelities of $99.61\pm0.04\%$ and $99.82\pm0.02\%$, respectively, as well as a series of diabatic CPhase gates with high fidelities characterized by cross-entropy benchmarking. The consistency and robustness of our technique are further validated by the reduction in pulse distortion and phase error observed across multilayer CZ gates. These results underscore the potential of our calibration and predistortion method to enhance the performance of two-qubit gates in superconducting quantum processors.

\end{abstract}

\maketitle


\section{\label{sec1}Introduction}

Superconducting quantum processors have emerged as a competitive platform for quantum computing, offering scalable architectures and high-fidelity operations. A key component in these processors is the tunable coupler~\cite{Niskanen2007, Yan2018}, which is capacitively coupled to two nearby qubits. The tunable coupler plays a crucial role in the scalability and functionality of superconducting quantum processors by enabling selective control over the interaction between qubits. Specifically, it allows for the adjustment of both transverse and longitudinal coupling, facilitating the isolation of qubits for single-qubit operations and the 
entanglement for two-qubit gate operations, such as controlled-phase gates (CPhase, CZ)~\cite{Collodo2020, Xu2020, Xu2021, Ye2021, Stehlik2021, GoogleQuantumAI2023, Simakov2023, Ding2023, Zhang2024, Barends2019a, Sung2021, Moskalenko2022} and fermionic simulation gates (fSim, iSWAP)~\cite{Arute2019, Barends2019a, Foxen2020, Han2020, Sete2021b, Sung2021, Moskalenko2022, Yun2024}.

The tunable coupler shares similar architecture with a conventional physical qubit~\cite{Niskanen2007}, which typically includes XY and Z control lines for qubit excitation and frequency modulation respectively, and a readout resonator. However, the effective coupling between qubits is primarily determined by the resonant frequency of the coupler~\cite{Yan2018}. Therefore, the excitation and readout of the coupler are often unnecessary, and these additional components can introduce hardware overhead, increase control complexity, and contribute to decoherence noise. To mitigate these issues, recent designs of superconducting quantum processors have simplified the coupler structure, retaining only the Z control line, to bias the coupler frequency by controlling the flux~\cite{Arute2019, Zhu2022, Ren2022, Cao2023, Xu2023, Yao2023, Xu2024a, Bao2024}.

High-precision control of the flux signal on the Z line is essential for achieving high-fidelity, stable, and repeatable two-qubit gates~\cite{Rol2019, Negirneac2021}. However, distortions in the input voltage signal on the Z line can accumulate over repeated gate operations, leading to phase or leakage errors. Various efforts have been made to address these distortions. Early approach involves using the coupler XY line and readout resonator to calibrate the Z pulse~\cite{Sung2021}, similar to methods used for qubit Z pulse calibration~\cite{Yan2019, Rol2019}. However, these additional structures have been phased out due to hardware complexity and increased noise. Another method involves exciting the coupler via crosstalk from nearby qubits and reading its state through the AC Stark effect using another qubit as a probe~\cite{Xu2020, Shi2023}. This approach faces the challenge with the limited strength of on-chip crosstalk effects, especially in microfabrication techniques like flip-chip~\cite{Foxen2018, Liang2023} and pogo pin packaging~\cite{Bronn2018}. The small qubit frequency shift, around 5 MHz, induced by the AC Stark effect also makes it difficult to achieve clear and reliable measurements~\cite{Zhang2023}. Furthermore, some researchers have attempted to detect the phase of nearby qubit after applying a flux square wave~\cite{Cao2023, Zhao2024, Li2024a}. However, this method may be limited by the presence of non-negligible dephasing noises on state-of-the-art superconducting quantum processors, leading to unstable and noisy results.

In this paper, we introduce a high-precision calibration method for the flux signals of tunbale couplers, which neither requires coupler excitation or readout, nor relies on measuring the nearby qubit phase. By leveraging the significant frequency shift of nearby qubit due to its strong transverse (XY) coupling with the coupler, we can directly probe the Z-pulse distortion by observing the population of the qubit excitation when it approaches resonance with the coupler. Unlike previous methods which calibrate the combination of two time-different step responses because a square wave contains two steps, we directly measure both short-time and long-time step responses of coupler Z transient with only one signal step. After predistorting the coupler Z pulses, we compare the distortion measured by other methods and evaluate the phase errors during repeated CZ gates. We finally achieve high-fidelity, repeatable, and stable two-qubit gates, including CZ, controlled-phase (CPhase), and iSWAP gates using our pulse calibration and predistortion method.

\section{\label{sec2}Experimental Systems}

\begin{figure}[!htb]
\includegraphics[scale=0.19]{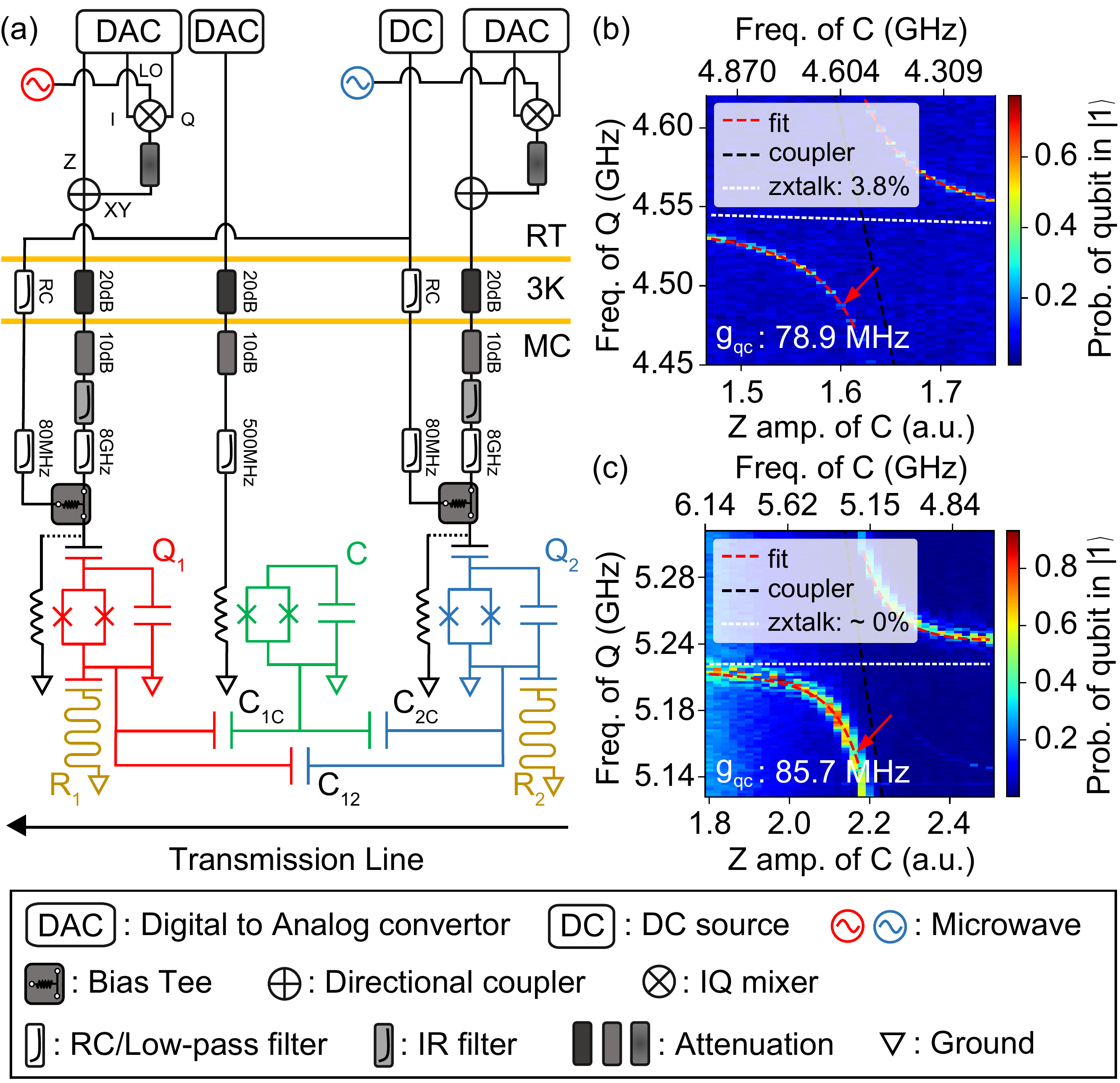}
\caption{\label{fig1}Experimental setup. (a) Wiring information of both chips. The legend below provides details for all icons. (b) and (c) Experimental data of anti-crossing between qubit and coupler for planar (Chip1) and flip-chip (Chip2) processors, respectively. The red dashed lines represent the fitted qubit frequency for anti-crossing, while the black dashed lines show the bared coupler spectrum. The qubit frequency changes with coupler Z amplitude, shown as white dashed lines, indicating the strength of the classical Z crosstalk effect. The red arrows point to the working points (near resonance) of our following experiments.}
\end{figure}

Consider a general tunable coupling system for three modes (Fig.~\ref{fig1}(a)), where two qubits ($Q_1$ and $Q_2$) are coupled to a tunable coupler ($C$) with $g_{ic}\ (i=1,2)$ being transverse coupling strength between $Q_i$ and $C$, and $g_{12}$ being the direct coupling strength between qubits. From the perspective of truncating to the subspace of one qubit and the coupler, the Hamiltonian stands as ($\hbar=1$)
\begin{equation}
    \label{equ1}
    H = -\sum_{i=q,c}\frac{1}{2}\omega_i\sigma_i^z + g_{qc} \left(\sigma_q^+\sigma_c^- + \sigma_q^-\sigma_c^+\right),
\end{equation}
where $\sigma^z_i$, $\sigma^-_i$, and $\sigma^+_i$ are the Pauli-Z, lowering, and raising operators for the qubit (q) and coupler (c), respectively. This strong coupling produces two dressed states $\ket{\phi^\pm}$, with eigenenergies
\begin{equation}
    \label{equ2}
    \omega^\pm=\frac{1}{2}\left[(\omega_q+\omega_c)\pm\Delta\right],
\end{equation}
where $\Delta^2=(\omega_q-\omega_c)^2+4g_{qc}^2$. The energy levels as a function of the coupler frequency are shown in Fig.~\ref{fig1}(b), where the minimum difference between the eigenenergies is $2g_{qc}$ when the system is in resonance ($\omega_q = \omega_c$). 

We use two chips in our experiments: one is a 20-qubit (19-coupler) planar superconducting processor (referred to as Chip1~\cite{Jin2024b, Jin2024a}) with $Q_{19}$, $Q_{20}$ and $C_{1920}$ employed in our study; another is a 21-qubit (20-coupler) flip-chip superconducting processor (referred to as Chip2) with $Q_{7}$, $Q_{8}$ and $C_{78}$ used. The constructional details and basic parameters of both processors are provided in Appendix~\ref{appA}.

We compare the XY and Z crosstalk between Chip1 and Chip2 (detailed in Appendix~\ref{appB}) to demonstrate that these crosstalk effects are predominantly restricted to the flip-chip processor (Chip2). Specifically, the planar processor (Chip1) exhibits non-negligible classical Z crosstalk, which slightly alters $\omega_q$ with different coupler Z amplitudes. To address this, we provide an improved method for fitting the anticrossing, as shown in Fig.~\ref{fig1}(b) and (c), which takes into account the linear Z crosstalk effect (detailed in Appendix~\ref{appB}). Furthermore, the low XY crosstalk observed on the flip-chip processor (Chip2) makes it challenging to excite the coupler. This highlights the importance of the method we propose for calibrating and correcting the Z pulse for the coupler, making it particularly useful in such scenarios.

\section{\label{sec3}Pulse calibration and correction of tunable couplers}

The system is initialized with the coupler near resonance with the qubit from a higher frequency, which means the qubit is always at a lower frequency, as the red arrow shown in Fig.~\ref{fig1}(b) and (c). Then, by applying a microwave field with an angular frequency of $\omega^-$ to the XY line of the qubit, the eigenstate $\ket{\phi^-}$ (referred to as the qubit for simplicity) will be excited. If distortions occur while controlling the Z line of the coupler, the dressed frequency of the qubit will deviate substantially from the calibrated microwave frequency $\omega^-$, resulting in suboptimal excitation relative to the initial operating point. This is a criterion for characterizing the amount of distortion. A more accurate explanation about exciting this dressed state $\ket{\phi^-}$ is shown in Appendix~\ref{appD}. In our experiment, we select the working point where the qubit frequency is repelled by the coupler by approximately 50 MHz, which is of the same magnitude as the transverse coupling strength $g_{qc}\approx$ 80 MHz. The frequency shift achieved by our method is significantly larger than that obtained using the AC Stark effect~\cite{Zhang2023}, ensuring a higher precision of the distortion calibration.

\begin{figure}[!htb]
\includegraphics[scale=0.19]{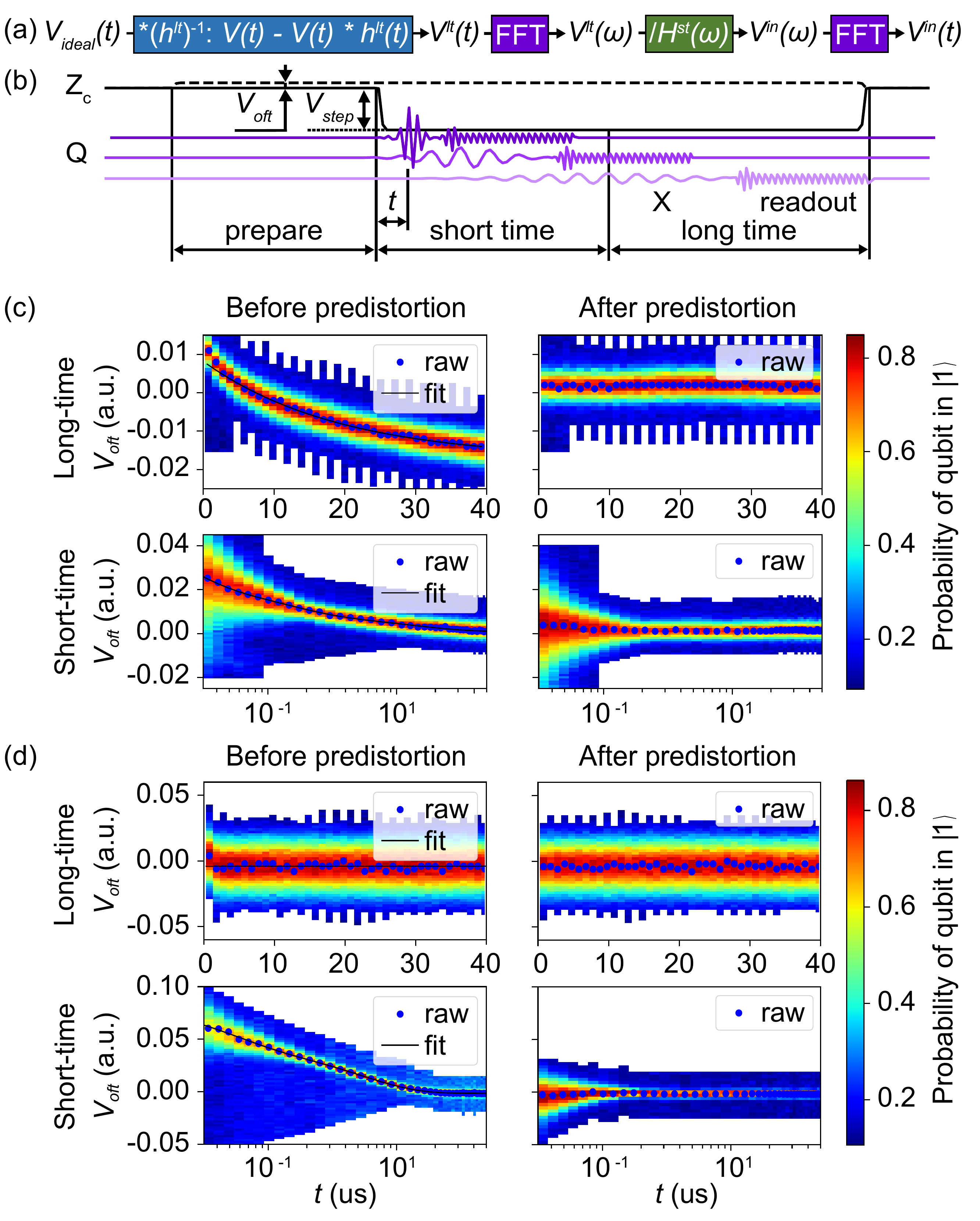}
\caption{\label{fig2}The calibration of coupler Z pulse. (a) The calibration and transformation procedure involved when generating a predistorted Z pulse. The correction of long-time distortion is performed using a second-order convolution as deconvolution in the time domain, after which the short-time distortion is corrected in the spectral domain. The pulse sequence for calibrating the coupler Z pulse is shown in (b). Experimental data of long-time and short-time distortion before and after predistortion are shown in (c) (Chip1) and (d) (Chip2). The black solid lines show the fitting results for the measured distortions.}
\end{figure}

Assuming the entire system is a linear and time-invariant (LTI) system~\cite{Butscher2018}, we can measure the step response of the Z pulse for the coupler, i.e. $s(t)$, and apply the following convolution to correct the input Z pulse $V_{\rm \mathit{in}}(t)$:
\begin{equation}
    V_{c}(t)=V_{\rm \mathit{in}}(t)\ast h(t)=\int_{-\infty}^{\infty} \mathrm{d}\tau~\!V_{\rm \mathit{step}}~\!u(t)~\!h(t-\tau),
\end{equation}
where $h(t)$ is system impulse response and $u(t)$ denotes the Heaviside step response. The relationship between $s(t)$ and $h(t)$ is $s(t)=u(t)\ast h(t)$, resulting in $h(t)=s'(t)$~\cite{Yan2019}. The procedure of predistortion and typical experimental data are depicted in Fig.~\ref{fig2}.

Here we emphasize the importance of correcting the long-time distortion since it significantly affects the repeatability of high-fidelity CZ gates~\cite{Sung2021, Rol2019}. The long-time distortions are shown in the upper two subfigures in Fig.~\ref{fig2}(c) and (d). The working frequency of the qubit (red arrows) mentioned in Fig.~\ref{fig1}(b) and (c) is calibrated when the coupler Z pulse is biased with an amplitude of $V_{\rm \mathit{step}}$ at the beginning of the preparation stage (last for approximately 6 $\mu$s). The pulse sequence for calibrating step response $s^{\rm \mathit{lt}}(t)$ is shown in Fig.~\ref{fig2}(b), where a single Z step with the amplitude $V_{\rm \mathit{step}}$ is applied at the end of the preparation stage. We excite the qubit after a delay time $t$ to measure the distortion. For the long-time distortion, we sweep this delay time from 0 to 40 ${\rm \mu}$s. An additional offset voltage $V_{\rm \mathit{oft}}$ is applied to the coupler Z line in advance to compensate for the distortion. Here, we use a fixed 200 ns $\pi$ pulse to excite the qubit for a better resolution. The voltage points $V^{\rm \mathit{lt}}(t)$ that maximize the population of the qubit will represent the amount of distortion. We then use the exponential function with a relaxation time in magnitude of ${\rm \mu s}$ to fit the measured distortion and obtain the corresponding step response $s^{\rm \mathit{lt}}(t)$. The correction is performed using the deconvolution method instead of fast Fourier transformation (FFT)~\cite{Rol2020}, because at this time scale (around 20-40 ${\rm \mu s}$), FFT introduces significant numerical errors that can result in undesirable jump points. We use a second-order reversed convolution to approximate deconvolution~\cite{Yan2019} (detailed in Appendix~\ref{appC}). This process generates a predistorted pulse $V_{c}^{\rm \mathit{lt}}(t)$, which is then applied to correct the short-time distortion.

For short-time distortion (within 5000 ns), we perform the correction at the spectral level using FFT~\cite{Rol2020}. We extract the step response $s^{\rm \mathit{st}}(t)$ for a short-time level using a similar sequence, differing only in that the length of $\pi$ pulse is time-varying. We simultaneously change the length $t_\pi$ of the $\pi$ pulse for qubit with time, using the relationship $A t_\pi=$ constant (where $A$ is the amplitude of $\pi$ pulse), from 30 to 200 ns (see the purple lines in Fig.~\ref{fig2}(b)), ensuring better temporal resolution for short-time distortion and frequency resolution for long-time distortion respectively. The inputted signal is the predistorted pulse $V_{c}^{\rm \mathit{lt}}(t)$, and then the step response $s^{\rm \mathit{st}}(t)$ is fitted using a multi-exponential function. The impulse response will be attained as $h_{\rm \mathit{st}}(t)=-\Dot{V}^{\rm \mathit{st}}(t)/V_{\rm \mathit{step}}$, where minus comes from compensation rather than distortion. After predistortion, we finally generate the Z pulse $V_{c}^{in}(t)$, which differs from the ideal Heaviside step response $u(t)$ within $\sim 1\%$ at both short-time and long-time levels.

The calibration and correction results of Z pulse for Chip1 and Chip2 are shown in Fig.~\ref{fig2}(c) and (d), respectively, and all the curve fitting functions and parameters involved are detailed in Appendix~\ref{appC}. We also verified our method by comparing it with simulation results (see Appendix~\ref{appD}) and other experimental methods, including excitation and readout of coupler via AC stark effect~\cite{Zhang2023}, phase-detection of qubit~\cite{Cao2023, Zhao2024, Li2024a} and qubit-coupler swapping) (see Appendix~\ref{appE}). These results show consistency, and prove the efficiency of our method.

\section{\label{sec4}High-fidelity two-qubit gates}

\begin{figure}[!htb]
\includegraphics[scale=0.19]{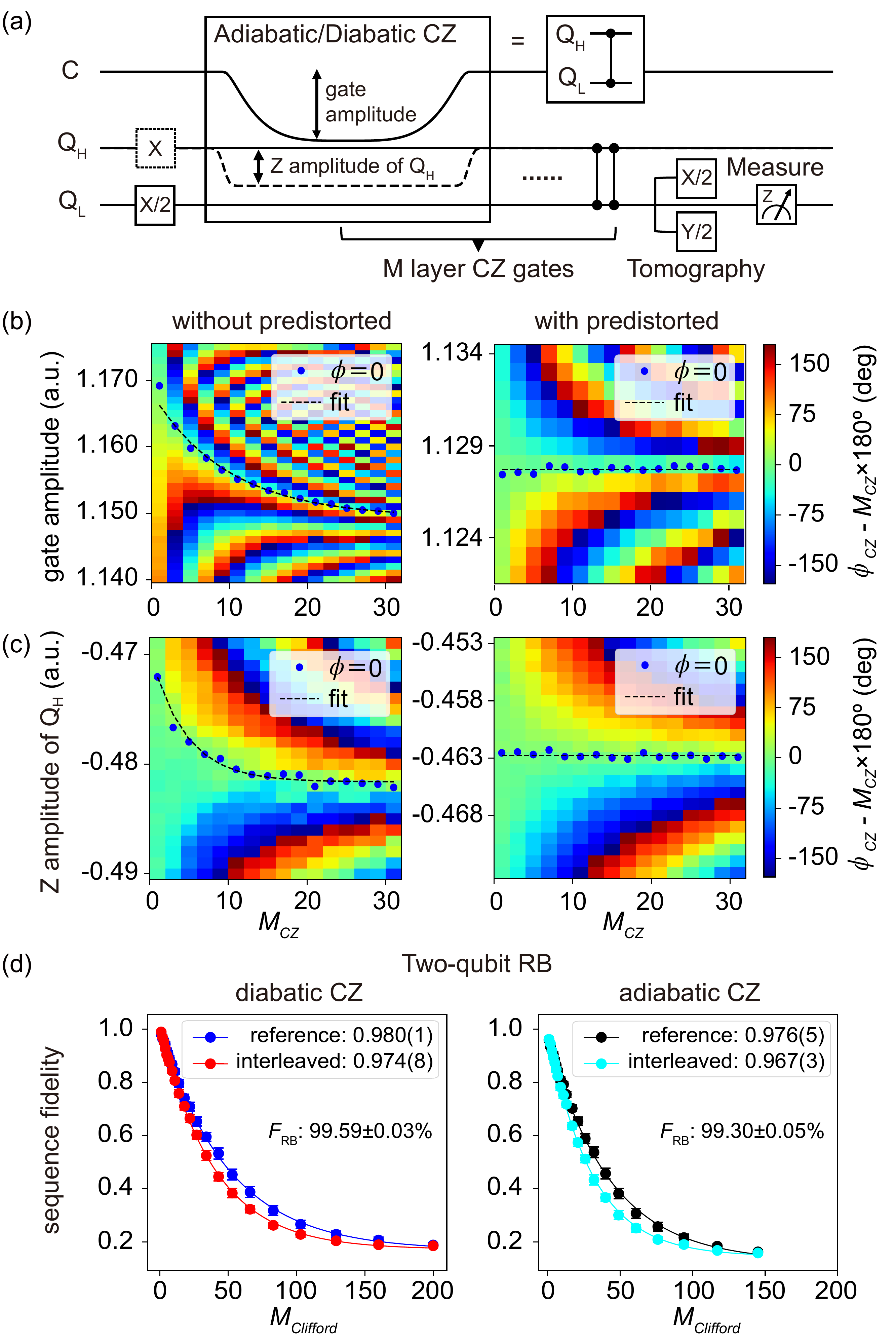}

\caption{\label{fig3}CZ gate performances with and without predistorted coupler Z pulse. (a) Pulse sequence for (b) adiabatic and (c) diabatic CZ gates. (b) and (c) Cross-Ramsey-like experiments to detect phase error ($\phi-N\pi$) by repeating CZ gates. The blue dots represent the zero phase error working points, and the black solid curves show the exponential fitting of these blue dots. (d) Interleaved randomized benchmarking fidelities. The error bars represent the standard deviation of $k=40$ random sequences.}
\end{figure}

To demonstrate the performance of Z pulse correction for the coupler, we show the experimental results of multilayer CZ gates with and without predistortion in Fig.~\ref{fig3}. The adiabatic controlled-Z (CZ) gate is implemented on Chip2 (illustrated in Fig.~\ref{fig3}(a)) through the adiabatic control of the coupler frequency, transitioning from an idle point (where the ZZ interaction between qubits is nearly negligible) to a working point (where the ZZ interaction is activated), leading to the accumulation of an additional phase on the computational state $\ket{11}$~\cite{Shi2023}. Note that only the coupler is biased via the fast Z pulse to ensure that the results demonstrate the performance of Z pulse correction.

To measure the conditional phase generated by the CZ gate, we perform a cross-Ramsey-like~\cite{Sung2021} experiment (Fig.~\ref{fig3}). The control qubit ($Q_{\rm H}$) is prepared in state $\ket{0}$ or $\ket{1}$ respectively, while the target qubit ($Q_{\rm L}$) is prepared in state $\ket{+}$ by using $\pi/2$ pulse. Then, $N$ repeated CZ gates are performed, after which single qubit quantum state tomography (QST) is applied to the target qubit. The phase difference of states for $Q_{\rm L}$ between $Q_{\rm H}$ initialized in $\ket{0}$ and $\ket{1}$ state is the conditional phase that computational state $\ket{11}$ accumulates. Ideally, the Z pulse amplitude of the coupler for repeated CZ gates which determines the accumulated $\pi$ phase should remain the same. Hence, the pattern of the cross-Ramsey-like experiment as a function of coupler Z pulse amplitude (vertical axis) and repeated CZ gates (horizontal axis) should represent a chevron-like shape. When the well-calibrated predistortion is applied (right of Fig.~\ref{fig3}(b)), the chevron-like shape is perfectly symmetric, and the repeated working point is obvious as the dashed line displayed. However, without predistortion, the working points differ with repeated CZ pulses, resulting in phase error (accumulating more or less than $\pi$ phase). The pattern shifts toward smaller coupler amplitudes with the increasing number of repeated CZ gates (dashed line, left of Fig.~\ref{fig3}(b)) indicating significant phase error ($\phi-N\pi$) originating from Z pulse distortion of the coupler.

The additional dynamic phases of qubits are measured through quantum process tomography (QPT) and eliminated through virtual Z gates~\cite{Krantz2019, SCPMA_Chen2022}. The gate fidelity is characterized using interleaved randomized benchmarking (IRB)~\cite{Magesan2012, Corcoles2013}, as shown in Fig.~\ref{fig3}(d). The result with predistorted Z pulse control of the coupler shows a high fidelity of $F_{\rm RB}=99.30\pm0.05\%$, where the uncertainty is calculated from the fitting errors (details are shown in Appendix~\ref{appG}). The distortion correction eliminates this phase error, enabling a higher fidelity, more repeatable, and stable CZ gate.

We also realize a diabatic CZ gate (Fig.~\ref{fig3}(a)) of higher fidelity on this flip-chip processor (Chip2) by biasing the coupler using the same waveform to shift the higher-level $\ket{2}$ of $Q_{\rm H}$ to interact with $\ket{1}$ of $Q_{\rm L}$. After the dynamic of exchanging population of state $\ket{11}$ with $\ket{20}$ for a whole period, computational state $\ket{11}$ accumulates $\pi$ phase compared with other computational states. Note that the accumulated $\pi$ phase is sensitive to the biasing Z pulse amplitude of $Q_{\rm H}$, instead of that of the coupler, so we deliver this similar cross-Ramsey-like experiment as a function of $Q_{\rm H}$ Z pulse amplitude and repeated CZ gates, as shown in Fig.~\ref{fig3}(c). The symmetry of the chevron-like pattern also recovers when the predistorted coupler Z pulse is applied (right in Fig.\ref{fig3}(c)). This diabatic CZ gate with a similar gate length (for controlling the influence of decoherence) shows a higher fidelity of $F_{\rm RB}=99.59\pm0.03\%$ (Fig.~\ref{fig3}(d)).

The fidelities of these two kinds of CZ gates are also measured using cross-entropy benchmarking (XEB)~\cite{Boixo2018, Arute2019} (detailed in Appendix~\ref{appF}). The average XEB fidelities of the adiabatic and diabatic CZ gates are $F_{\rm XEB}=99.37\pm0.04\%$ and $F_{\rm XEB}=99.61\pm0.04\%$, respectively, which are consistent with the average RB fidelities.

\begin{figure}[!htb]
\includegraphics[scale=0.19]{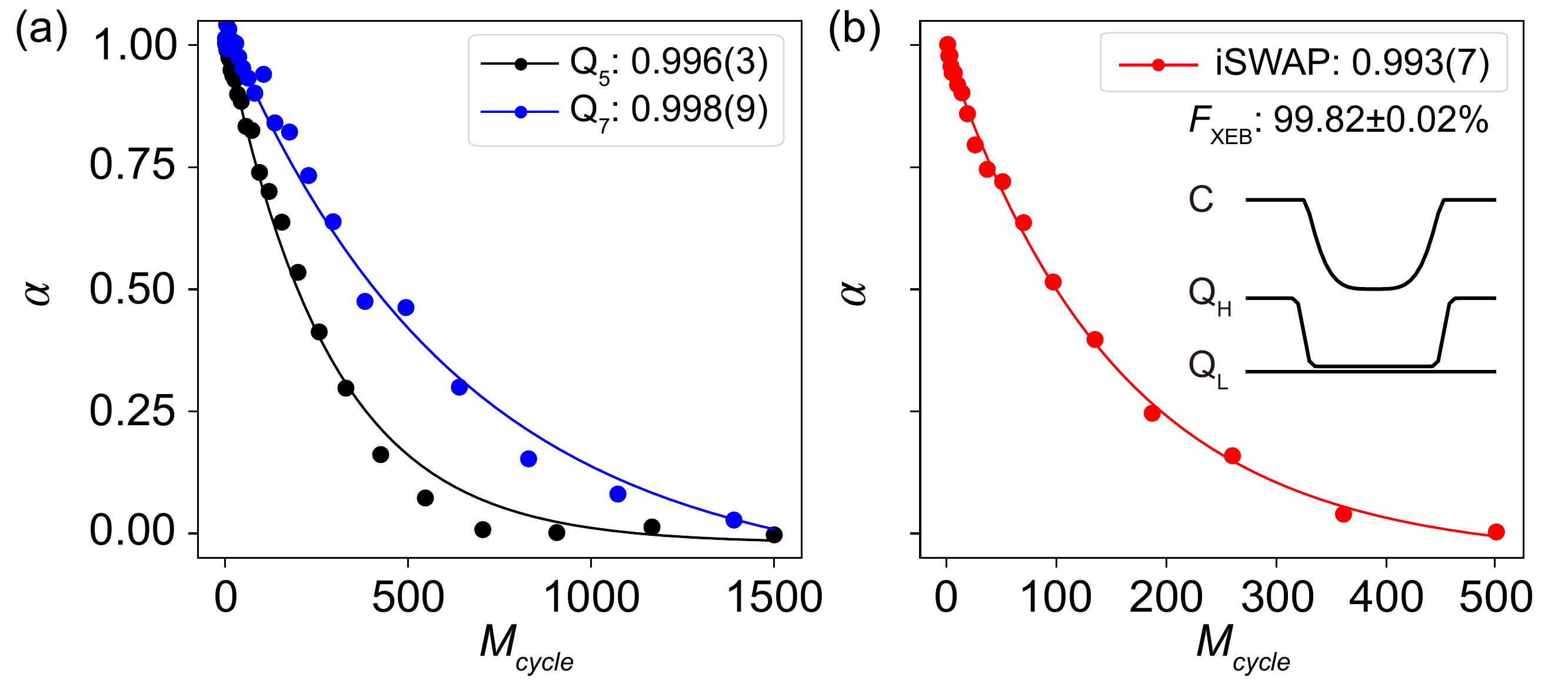}

\caption{\label{fig4}Cross-entropy benchmarking for (a) single-qubit and (b) iSWAP gates. Fidelities $\alpha$ are obtained from $k=50$ random sequences. The inset of (b) shows the pulse sequence of the iSWAP gate.}
\end{figure}

In addition, we demonstrate the effectiveness of our method by implementing other two-qubit gates, including an iSWAP gate and a series of CPhase gates with different conditional phases. In Fig.~\ref{fig4}, we demonstrate a competitive fidelity iSWAP gate through the coherent exchange. The gate amplitude of the coupler is carefully selected to suppress the residue ZZ interactions~\cite{Sung2021}, resulting in a near-zero conditional phase. Fidelity results from XEB of single qubits and interleaved iSWAP gate cycles are shown in Fig.~\ref{fig4}(a) and (b), respectively. The average XEB fidelities of the diabatic iSWAP gate is $F_{\rm XEB}=99.82\pm0.02\%$. In Appendix~\ref{appF}, we also detail a series of CPhase gates with $\pi/2$, $\pi/4$ and $\pi/8$ conditional phases using diabatic coherent exchange. The corresponding XEB fidelities are $99.71\pm0.03\%$, $99.81\pm0.03\%$ and $99.94\pm0.04\%$, respectively.

\section{\label{sec5}Conclusion}
In this paper, we present a method to calibrate both short-time and long-time distortions of the Z pulse transient for the coupler with enhanced resolution, surpassing that of prior techniques. Our approach achieves this calibration without exciting the coupler, employing a dedicated readout resonator for state measurement, or assessing the phase of an adjacent qubit. A key advantage of our method lies in its direct calibration of the step response following a single Z step, as opposed to the conventional approach of measuring the Z distortion, which encompasses two distinct step responses over time.

We further introduce a scheme for calculating predistortion, which addresses short-time distortions by applying fast Fourier transformation in the spectral domain and corrects long-time distortions through second-order convolution as a deconvolution technique in the time domain. Utilizing the predistorted Z pulse, which exhibits an error margin of approximately $1\%$, we conduct a cross-Ramsey-like experiment to quantify the cumulative phase error across repeated CZ gates, both with and without predistortion, thereby illustrating the effectiveness of our correction method.

In addition, we successfully calibrate the adiabatic and diabatic CZ gates, a series of CPhase gates featuring conditional phases of $\pi/2$, $\pi/4$, and $\pi/8$, as well as an iSWAP gate on the flip-chip processor (Chip2). These operations are characterized by high fidelity (refer to Table~\ref{table1}), repeatability, and stability, highlighting the robustness and versatility of our approach. Our work offers a simpler and more robust solution for precise flux control of tunable couplers. By reducing the complexity and improving the reliability of the calibration process, our method paves the way for more efficient and scalable quantum computing systems for large-scale implementation of high-fidelity quantum two-qubit gates.


\begin{table}[!htb]
\caption{\label{table1}Conditional phase $\phi$, swap angle $\theta$, and average XEB fidelities for two-qubit gates.}
\begin{ruledtabular}
\begin{tabular}{cccc}
Gate & $\phi$ (deg) & $\theta$ (deg) & XEB fidelity ($\%$)\\
\hline
CZ (adiabatic)      & 180  & 0  & $99.37\pm0.04$ \\
CZ (diabatic)       & 180  & 0  & $99.61\pm0.04$ \\
CPhase ($\pi/2$)    & 90   & 0  & $99.71\pm0.03$ \\
CPhase ($\pi/4$)    & 45   & 0  & $99.81\pm0.03$ \\
CPhase ($\pi/8$)    & 22.5 & 0  & $99.94\pm0.04$ \\ 
iSWAP               & 0    & 90 & $99.82\pm0.02$ \\
\end{tabular}
\end{ruledtabular}
\end{table}

This method can also calibrate the distortion of the flux control lines for any other system, where there is a strong transverse coupling between it and the probe, such as the central bus resonator~\cite{McKay2016, Guo2021b}. Its specialty is particularly evident in scenarios where it is unnecessary and difficult to excite the mode of this system but necessary to bias the frequency.

\begin{acknowledgments}
We acknowledge the support from the Synergetic Extreme Condition User Facility (SECUF) and the Quafu Quantum Cloud Computing Cluster (https://quafu.baqis.ac.cn/). This work was supported by the National Natural Science Foundation of China (Grants Nos. T2121001, 92265207, T2322030, 12122504, 12274142, 92365206, 12104055), the Innovation Program for Quantum Science and Technology (Grant No. 2021ZD0301800), the Beijing Nova Program (No. 20220484121), and the China Postdoctoral Science Foundation (Grant No. GZB20240815). 
\end{acknowledgments}

\appendix

\renewcommand{\thefigure}{\Alph{section}\arabic{figure}}
\renewcommand{\thetable}{\Alph{section}\arabic{table}}

\setcounter{figure}{0}
\setcounter{table}{0}

\section{Devices parameters\label{appA}}

\begin{figure*}[!htb]
\includegraphics[scale=0.4]{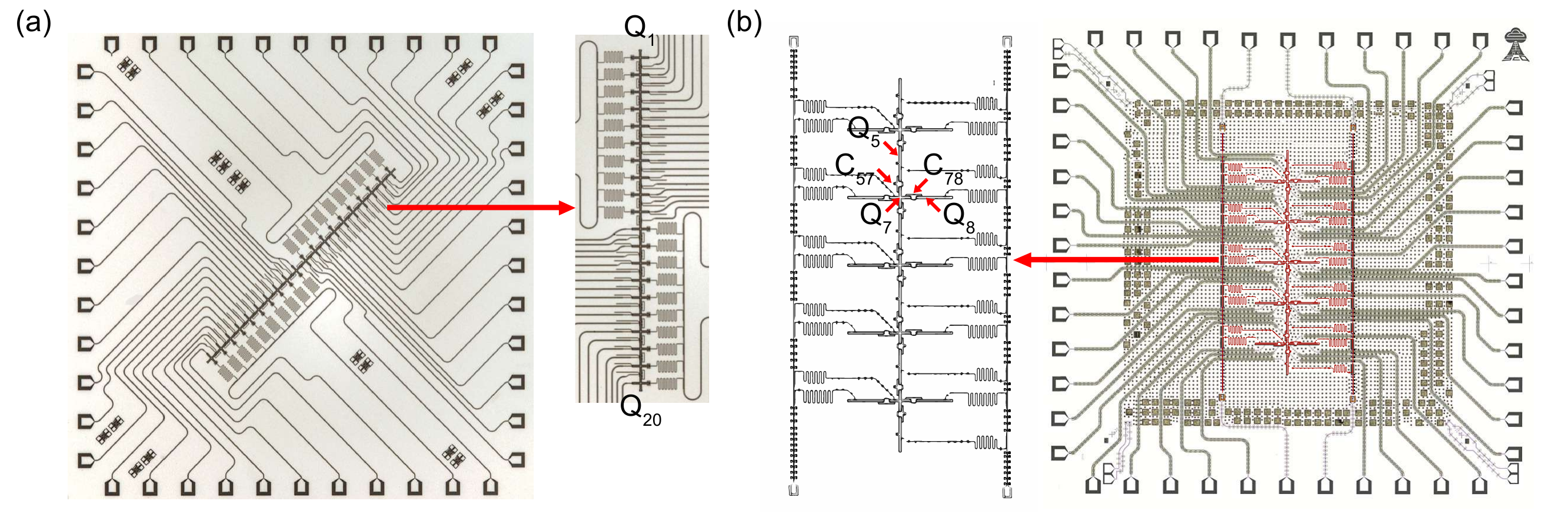}
\caption{\label{figA1}Quantum processors. (a) Planar processor (Chip1). Qubits ($Q_1$-$Q_{20}$) are arrayed in a row. (b) Flip-chip processor (Chip2). The red layer represents the upper chip layer. Qubits ($Q_1$-$Q_{21}$) are arrayed as a 1D chain with multiple legs.}
\end{figure*}

For clarity, all the couplers are labeled according to the numbers of the qubits they connect.

\textbf{Chip1:} a planar superconducting processor consists of 20 transmon qubits ($Q_1$-$Q_{20}$) and 19 couplers ($C_1$-$C_{19}$) arrayed in a row (shown in Fig.~\ref{figA1}(a)). We focus on QCQ pair $Q_{19}$-$C_{1920}$-$Q_{20}$, and their basic parameters are listed in Tab.~\ref{tableA2}.

\textbf{Chip2:} a flip-chip superconducting processor consists of 21 transmon qubits ($Q_1$-$Q_{21}$) and 20 couplers ($C_{12}$-$C_{1821}$) arrayed with a 1D chain with multiple legs (shown in Fig.~\ref{figA1}(b)). We focus on QCQ pairs $Q_5$-$C_{57}$-$Q_7$ and $Q_7$-$C_{78}$-$Q_8$, and their basic parameters are listed in Tab.~\ref{tableA1}.

The Josephson junctions in the superconducting quantum interference device (SQUID) of the couplers display either symmetrical or asymmetrical configurations in Chip1 or Chip2. The design of Chip2 specifically enhances the stability of flux control over the coupler frequency.

Our Chip2 (bottom) is depicted in Fig.~\ref{figA1}(b). The device was fabricated on two sapphire substrates: one measuring 11$\times$11 ${\rm mm}^2$ (top) and the other 15$\times$15 ${\rm mm}^2$ (bottom), each with a thickness of 430 $\mu$m. Initially, a 100 nm thick Al base layer was deposited across the entire substrate surface via electron-beam evaporation and patterned photolithographically. The base-layer structures, which include transmission lines, microwave coplanar waveguide resonators, XYZ control lines, and qubit capacitors, were defined using a wet etching process. The Josephson junctions were fabricated using a double-angle evaporation technique. The undercut structure was prepared using a PMMA-MMA double-layer resist, following a process similar to that reported in Ref.~\cite{Barends2013}. A 65 nm thick Al bottom electrode was evaporated at a 60\degree\ angle, followed by oxidation in pure oxygen. Then, a 100 nm thick Al counter electrode was deposited at a 0\degree\ angle. After lift-off, airbridges~\cite{Chen2014a} were fabricated to suppress parasitic modes and reduce crosstalk between control lines. Finally, laser direct writing was employed to expose the indium bumps, which serve to connect the top and bottom chips. These indium bumps, approximately 7.5 $\mu$m in height, were grown using an electron beam evaporation device. Flip-chip bonding of the two chips was then carried out using FC150 equipment. The fabrication of Chip1 followed a similar process, with additional details discussed in the previous work~\cite{Guo2021c}.

\begin{table}[!htb]
\caption{\label{tableA1}Device parameters for $Q_{19}$-$C_{1920}$-$Q_{20}$ on planar superconducting processor (Chip1).}
\begin{ruledtabular}
\begin{tabular}{cccc}
 & $Q_{19}$ & $C_{1920}$ & $Q_{20}$\\
\hline
$\omega/2\pi$\footnotemark[1] (GHz)
& 4.5780 & 5.7405 & 4.9040 \\
$\omega_{\rm \mathit{sw}}/2\pi$\footnotemark[2] (GHz)
& 4.6947 & 6.3031 & 5.1886 \\
$\eta/2\pi$\footnotemark[3] (MHz)
& 203 & 299 & 196 \\
$T_1$\footnotemark[4] ($\mu {\rm s}$)
& 27.9 & & 37.4 \\ 
$T_2$\footnotemark[4] ($\mu {\rm s}$)
& 2.3 & & 3.0 \\ 
$g/2\pi$\footnotemark[5] (MHz)
& 78.9 & 10.0 & 89.8 \\
$\omega_r/2\pi$\footnotemark[6] (GHz)
& 6.7874 & & 6.8069 \\
$F_0$\footnotemark[7] (\%)
& 94.3 & & 95.1 \\
$F_1$\footnotemark[7] (\%)
& 89.8 & & 87.7 \\
\end{tabular}
\end{ruledtabular}
\footnotetext[1]{$\ket{0}\rightarrow\ket{1}$ transition frequencies at the idle working point.}
\footnotetext[2]{Sweet-point transition frequencies.}
\footnotetext[3]{Anharmonicities at the idle working point.}
\footnotetext[4]{Energy decay time ($T_1$), Ramsey decay time ($T_2$) measured at the idle working point.}
\footnotetext[5]{Coupling strengths. $g_{1c}$, $g_{12}$ and $g_{2c}$, respectively from left to right.}
\footnotetext[6]{Readout resonator frequency.}
\footnotetext[7]{Readout fidelities $F_0$($F_1$) for preparing $\ket{0}$($\ket{1}$) and reading as $\ket{0}$($\ket{1}$).}
\end{table}

\begin{table}[!htb]
\caption{\label{tableA2}Device parameters for $Q_5$-$C_{57}$-$Q_7$-$C_{78}$-$Q_8$ on flip-chip superconducting processor (Chip2).}
\begin{ruledtabular}
\begin{tabular}{cccccc}
 & $Q_5$ & $C_{57}$ &$Q_7$ & $C_{78}$ & $Q_8$\\
\hline
$\omega/2\pi$ (GHz)
& 4.9830 & 8.2850 & 4.9030 & 7.6635 & 5.1850 \\
$\omega_{\rm \mathit{sw}}/2\pi$ (GHz)
& 5.0539 & 7.4520 & 4.9103 & 8.3157 & 5.1852 \\
$\eta/2\pi$ (MHz)
& 200 & 298 & 200 & 300 & 205 \\
$T_1$ ($\mu {\rm s}$)
& 15.3 & &35.0 & & 47.9 \\ 
$T_2$ ($\mu {\rm s}$)
& 4.5 & &11.6 & & 20.1 \\ 
$g/2\pi$ (MHz)
& 81.8 & 2.1 & 83.4 & 2.2 & 85.7 \\
$\omega_r/2\pi$ (GHz)
& 6.9211& &6.8627 & & 6.8297 \\
$F_0$ (\%)
& 99.1& &93.3 & & 98.3 \\
$F_1$ (\%)
& 87.5& &83.0 & & 92.5 \\
\end{tabular}
\end{ruledtabular}
\end{table}

\section{Crosstalk effects and anti-crossing\label{appB}}
\setcounter{figure}{0}
\setcounter{table}{0}

Since the control lines and the circuits (qubits and couplers) of the planar processor (Chip1) are on the same plane, crosstalk effects are significant for one qubit (coupler) with its nearby control lines for another qubit (coupler). However, for the flip-chip processor (Chip2), several factors contribute to the alleviation of crosstalk. Firstly, the lines are designed to be more widely spaced. Secondly, the lines and circuits are on different planes. Lastly, these planes are connected by special structures called indium bumps~\cite{Foxen2018}, which serve as Faraday cages to eliminate the propagation of electromagnetic fields. The Z line crosstalk strength is detailed in Tab.~\ref{tableB1}.

\begin{table}[!htb]
\caption{\label{tableB1}Z line crosstalk strength. Add Z pulse on $Q$/$C$ and check frequency change of another $Q$.}
\begin{ruledtabular}
\begin{tabular}{ccccc}
line $\rightarrow Q$ 
& $Q_7\rightarrow Q_8$ & $Q_8\rightarrow Q_7$
& $C_{78}\rightarrow Q_7$ & $C_{78}\rightarrow Q_8$\\
\hline
Strength & 0.16\% & 0.08\% & \textless 0.01\% & \textless 0.01\%\\
\hline
\hline
line $\rightarrow Q$ 
& $Q_{19}\rightarrow Q_{20}$ & $Q_{20}\rightarrow Q_{19}$
& $C_{1920}\rightarrow Q_{19}$ & $C_{1920}\rightarrow Q_{20}$\\
\hline
Strength & 0.61\% & 0.38\% & 3.75\% & 0.03\%\\
\end{tabular}
\end{ruledtabular}
\end{table}

For Chip1, classical Z crosstalk can be mainly eliminated using a simple linear model, which adds a square wave with an amplitude related to the strength of the crosstalk. However, accurately determining the linear coefficient is challenging due to anti-crossing phenomena or other nonlinear quantum effects. Therefore, we provide a method for fitting anti-crossing, which enables us to gain the crosstalk coefficient or correct the imperfect coefficient and achieve a more stable coupling strength simultaneously.

At a crosstalk level of around $3\%$ or below, the qubit frequency is linearly affected by the Z pulse amplitude of the coupler due to Z crosstalk, as described by the equations
\begin{equation}
    f_q = k_q {\rm zpa}_q + b_q
\end{equation}
and
\begin{align}
    f_q({\rm zpa}_c) 
    &= k_q\left({\rm zpa}_c + {\rm coeff}_{\rm zxtalk}\cdot {\rm zpa}_c\right) + b_q\\
    &= k_{\rm \mathit{eff}}{\rm zpa}_c + b_{\rm \mathit{eff}}
\end{align}
where ${\rm coeff}_{\rm zxtalk}$ is the coefficient modeling the crosstalk effect as a linearly scaled Z pulse, $f_q=2\pi\omega_q$ represents the idle frequency of qubit, and ${\rm zpa}_q$ is the Z amplitude of qubit at its idle point. $k_q$ and $b_q$ describe this linear relationship, where $k_q$ can be approximately estimated from qubit's spectroscopy near its idle point as:
\begin{equation}
    \label{equB4}
    k_q = \frac{f_q(\max({\rm zpa}_c))-f_q(\min({\rm zpa}_c))}{{\rm coeff}_{\rm zxtalk}(\max({\rm zpa}_c)-\min({\rm zpa}_c))}.
\end{equation}

\begin{figure}[!htb]
\includegraphics[scale=0.19]{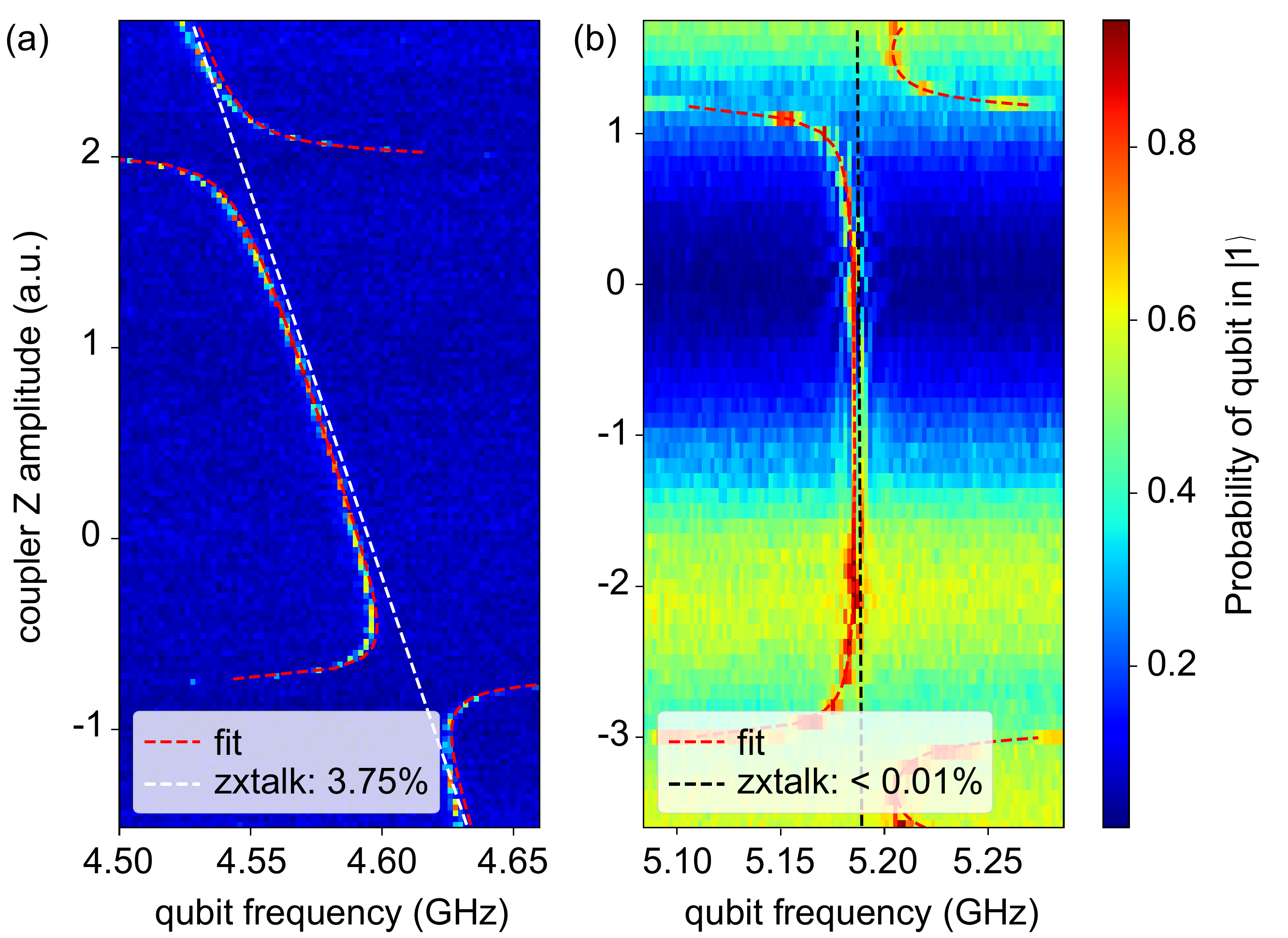}
\caption{\label{figB1}Typical experimental data of anti-crossing in $Q$-$C$ system. (a) Anti-crossing of $Q_{19}$-$C_{1920}$ on Chip1 (planar). (b) Anti-crossing of $Q_8$-$C_{78}$ on Chip2 (flip-chip). The red dashed curves represent the fitting of anti-crossing. The white dashed lines represent the fitted linear Z crosstalk strength from the coupler Z line to the qubit.}
\end{figure}

The function of anti-crossing is described by Eq.~\eqref{equ2}. By squaring the $\pm\Delta$ part, Eq.~\eqref{equ2} transforms into
\begin{equation}
    (f - f_q)(f - f_c) = g_{qc}^2,
\end{equation}
which provides an identity for coupling strength $g_{qc}$. Thus, the standard deviation (std) of $(f - f_q)(f - f_c)$ serves as a more stable error function for fitting anti-crossing and determining $g_{qc}$ from optimization. The overall perspectives of Fig.~\ref{fig1}(b) and (c) are shown in Fig.~\ref{figB1}. The linear crosstalk effect (red dotted line) is evident when ignoring the anti-crossing parts and is fitted to gain $k_{\rm \mathit{eff}}$. Using the relation $k_{\rm \mathit{eff}}=k_q{\rm coeff}_{\rm zxtalk}$ and Eq.~\eqref{equB4}, the crosstalk coefficient can be estimated.

Meanwhile, if the experimental data of anti-crossing is obtained with the correction of Z pulse crosstalk, this method can also check and correct the imperfect crosstalk coefficient. It is also useful for the flip-chip processor (Chip2) where the Z crosstalk effect is extremely faint (see Fig.~\ref{figB1}(b)).

Furthermore, we test the XY line crosstalk on both Chip1 and Chip2 by comparing the product of the $\pi$ pulse amplitude and length between situations of directly exciting the qubit using its XY line and exciting its nearby coupler using the same XY line of this qubit (shown in Tab.~\ref{tableB2}). It shows that exciting the coupler using the XY crosstalk effect is around 10 times more difficult on the flip-chip processor (Chip2) than on the planar processor (Chip1).

\begin{table}[!htb]
\caption{\label{tableB2}XY line crosstalk strength. Comparison of the strength of exciting $Q$ or $C$ using the same XY-line of $Q$.}
\begin{ruledtabular}
\begin{tabular}{cccc}
line $\rightarrow Q$/$C$
& $Q_8\rightarrow Q_8$ & $Q_8\rightarrow C_{78}$ & Ratio $C$/$Q$\\
\hline
Amplitude (a.u.) & 0.88 & 1.12 & \multirow{2}*{21.21}\\
Length (ns) & 30 & 500\\
\hline
\hline
line $\rightarrow Q$/$C$ 
& $Q_{19}\rightarrow Q_{19}$ & $Q_{19}\rightarrow C_{1920}$ & Ratio $C$/$Q$\\
\hline
Amplitude (a.u.) & 2.20 & 0.81 & \multirow{2}*{2.45}\\
Length (ns) & 30 & 200\\
\end{tabular}
\end{ruledtabular}
\end{table}

\section{The fitting function parameters and correction schemes for both short- and long-time distortion\label{appC}}
\setcounter{figure}{0}
\setcounter{table}{0}

Due to on-chip or on-control-line parasitic capacitance and inductance, the distortion of the Z pulse of the coupler exhibits exponential relaxation over several relaxation times. For our chip and control system, the short-time (under 5000 ns) distortion can be described as a superposition of multiple exponential functions, as follows:

\begin{equation}
\label{equC1}
    s_{\rm \mathit{\rm \mathit{st}}}(t) = \sum_{i=1,\cdots}^{\rm exp\_num}p_{i}e^{-t/\tau_{i}},
\end{equation}
where $\{\tau_i\}$ are the relaxation times for different exponential functions, and $\{p_i\}$ represent the relaxation amplitudes of these exponential functions.

Furthermore, the long-time (around 20-40 $\mu$s) distortion can be described using an exponential function, as follows:
\begin{equation}
\label{equC2}
    s_{\rm \mathit{lt}}(t) = u(t)\times \left((B-A)e^{-t/\tau}+A\right),
\end{equation}
where $A$ and $B$ are the coefficients related to the distortion and relaxation amplitude of the voltage signal for the coupler. 

All fitting function parameters for short-time and long-time distortion are detailed in Tab.~\ref{tableC1} and Tab.~\ref{tableC2}, respectively. The calibrations are applied using different methods: fast Fourier transformation (FFT) for the short-time response, and deconvolution for the long-time response. 

Deconvolution is realized by a second-order reversed convolution (2$^{\rm nd}$ RC)~\cite{Yan2019}. Let $X(j\omega)$ and $Y(j\omega)$ be the spectral domain of the predistorted input and ideal output signal, respectively. The relation between them is:
\begin{align}
    X(j\omega) 
    &= \frac{Y(j\omega)}{H(j\omega)} = \frac{Y(j\omega))}{1-R(j\omega)}\nonumber\\
    &= Y(j\omega)\left(1+R(j\omega)+R^2(j\omega)+\cdots\right)
    \label{equC3}
\end{align}
where $R(j\omega)=1-H(j\omega)$ represents the relative small distortion. The first term of Eq.~\eqref{equC3} is the ideal output signal, so the zero-order RC signal is $x_0(t)=y(t)$. The second term is the first perturbation term. After the Inverse Fourier transform (IFT), the time-domain relation is 
\begin{equation}
    x_1(t) = y(t) - y(t)\ast h(t).
\end{equation}
The third term is the second perturbation term. After IFT, the time-domain relation using the first-order RC signal $x_1(t)$ is 
\begin{equation}
    x_2(t) = x_1(t) - x_1(t)\ast h(t).
\end{equation}
The remaining terms are at least $R^3$ and are therefore ignored. Finally, we achieve the second-order RC signal $x_2(t)$ as the deconvolution of the input signal $y(t)$ for the long-time correction. This process effectively corrects the long-time distortion by iteratively reducing errors introduced by the transfer function of the system.

\begin{table}[!htb]
\caption{\label{tableC1}Fitting function parameters for short-time distortion on $C_{1920}$ (Chip1) and $C_{78}$ (Chip2).}
\begin{ruledtabular}
\begin{tabular}{ccc}
 & $C_{1920}$ & $C_{78}$\\
\hline
$p_i$
& -0.024, -0.011, -0.006 & -0.019, -0.021\\
$\tau_i$ (ns)
& 17.61, 132.07, 1305.15 & 47.83, 528.10\\
\end{tabular}
\end{ruledtabular}
\end{table}

\begin{table}[!htb]
\caption{\label{tableC2}Fitting function parameters for long-time distortion on $C_{1920}$ (Chip1) and $C_{78}$ (Chip2).}
\begin{ruledtabular}
\begin{tabular}{cccc}
 & $A$ & $B$ & $\tau$ (us)\\
\hline
$C_{1920}$
& 1.0127 & 0.9935 & 18.684\\
$C_{78}$
& \multicolumn{3}{c}{No long-time distortion}\\
\end{tabular}
\end{ruledtabular}
\end{table}

\section{Exciting the dressed state and numerical simulation of distortion\label{appD}}
\setcounter{figure}{0}
\setcounter{table}{0}

As mentioned in the main context, the Hamiltonian of the $Q$-$C$ system is given by Eq.~\eqref{equ1}. We focus on the single excitation manifold and shift the energy of the Hamiltonian so that the eigenenergy of state $\ket{00}$ for both the qubit and the coupler is zero. In the subspace of $\ket{01}$ and $\ket{10}$, $H_{sub}$ is given by
\begin{equation}
    \left[\begin{array}{cc}
    \omega_q & g_{qc} \\
    g_{qc} & \omega_c
    \end{array}\right],
\end{equation}
where the two states are coupled, and the coupling is significant when $\omega_q$ is close to $\omega_c$. After diagonalizing $H_{sub}$, we obtain two dressed eigenstates $|\phi^-\rangle$ and $|\phi^+\rangle$ and their eigenenergies as follows:
\begin{equation}
    \omega^\pm= \frac{1}{2} (\omega_c + \omega_q \pm \Delta)
\end{equation}
\begin{equation}
    \ket{\phi^-} = \frac{1}{A_1}\left(-\frac{\omega_c - \omega_q + \Delta}{2g_{qc}} \ket{10} + \ket{01}\right)
\end{equation}
\begin{equation}
   |\phi^+\rangle =  \frac{1}{A_2}\left(-\frac{\omega_c - \omega_q - \Delta}{2g_{qc}} \ket{10} + \ket{01}\right),
\end{equation}
where $A_1$ and $A_2$ are normalization factors for $\ket{\phi^-}$ and $\ket{\phi^+}$, respectively. Now we can rewrite $H_{sub}$ with $\ket{\phi^-}$ and $\ket{\phi^+}$:
\begin{equation}
    H_{sub} =\omega^{-} \ket{\phi^-}\bra{\phi^-} + \omega^{+} \ket{\phi^+}\bra{\phi^+}.
\end{equation}
Thus, no coupling exists between $\ket{\phi^-}$ and $\ket{\phi^+}$. In our calibration process, we need to excite the coupled system using only the XY-line of the qubit. Under the rotating wave approximation (RWA), the driving Hamiltonian can be written as:
\begin{equation}
    H_{d} = -\frac{\Omega}{2} \left(e^{\mathrm{i}\omega_dt}\ket{00}\bra{10} + e^{-\mathrm{i}\omega_dt}\ket{10}\bra{00}\right)
\end{equation}
where $\Omega$ is Rabi frequency, and $\omega_d$ is driving frequency.

Generally, this driving term will lead to the Rabi oscillation between $\ket{0}$ and $\ket{1}$ of the qubit if $\omega_d=\omega_q$. However, due to the strong coupling with the coupler, $Q$ and $C$ cannot avoid being influenced during the excitation process. Therefore, it is better to choose $\ket{\phi^-}$ and $\ket{\phi^+}$ as the states to describe the excitation process. To determine the influence of the driving Hamiltonian on the coupled system's evolution, we calculate the coupling matrix elements between $\ket{00}$ and $\ket{\phi^-}$($\ket{\phi^+}$),
\begin{equation}
    \bra{00}H_d\ket{\phi^+} =-\frac{\Omega}{2}e^{\mathrm{i}\omega_dt}\frac{\omega_c - \omega_q - \Delta}{2A_2g_{qc}}
\end{equation}
\begin{equation}
  \bra{00}H_d\ket{\phi^-} =-\frac{\Omega}{2}e^{\mathrm{i}\omega_dt}\frac{\omega_c - \omega_q + \Delta}{2A_1g_{qc}},
\end{equation}
from which we extract two effective Rabi frequencies, where $\Omega^+ = \Omega (\omega_c - \omega_q - \Delta)/(2A_2g_{qc})$ and $\Omega^- = \Omega (\omega_c - \omega_q + \Delta)/(2A_1g_{qc})$.

For the excitation process, neglecting leakage to higher energy levels, we consider only the subspace spanned by $\ket{00}$ and single-excitation states($\ket{\phi^-}$, $\ket{\phi^+}$). Then the Hamiltonian containing the driving Hamiltonian can be written as
\begin{align}
    H &= \omega^{-} \ket{\phi^-}\bra{\phi^-} + \omega^{+}\ket{\phi^+}\bra{\phi^+}\nonumber\\
      &-\frac{\Omega^-}{2}\left(e^{\mathrm{i}\omega_dt}\ket{00}\bra{\phi^-} + e^{-\mathrm{i}\omega_dt}\ket{\phi^-}\bra{00}\right)\\
      &-\frac{\Omega^+}{2}\left(e^{\mathrm{i}\omega_dt}\ket{00}\bra{\phi^+} + e^{-\mathrm{i}\omega_dt}\ket{\phi^+}\bra{00}\right).\nonumber
\end{align}
$|00\rangle$ interacts simultaneously with both $|\phi^-\rangle$ and $|\phi^+\rangle$, making the excitation process complex. However, if we set $\omega_d = \omega^-$, the interaction induced by the driving Hamiltonian can be simplified. We first move into a frame rotating with $\ket{\phi^-}$ at frequency $\omega^-$. Define $H_0=\omega^{-} \ket{\phi^-}\bra{\phi^-}$. For any state $\ket{\phi}$ that evolves according to the Hamiltonian $H$ during the excitation process, we define $\ket{\phi} = U_{H_0} \ket{\phi_{\rm \mathit{rf}}}$ , where $U_{H_0}$ is the propagator corresponding to $H_0$. To this end, we define
\begin{equation}
U_{\rm \mathit{rf}}=e^{\mathrm{i}H_0t}=U_{H_0}^\dagger
\end{equation}
and the new state in the rotating frame is $|\phi_{\rm \mathit{rf}}\rangle=U_{\rm \mathit{rf}}\ket{\phi}$. The time evolution in this new frame is again found from the Schrodinger equation: 
\begin{align}
    \mathrm{i} \frac{d}{dt}\ket{\phi_{\rm \mathit{rf}}}
    &= \mathrm{i} \left(\frac{d}{dt}U_{\rm \mathit{rf}}\right)\ket{\phi} + \mathrm{i} U_{\rm \mathit{rf}}\frac{d}{dt}\ket{\phi}\\\nonumber
    &= \mathrm{i} \left(\left(\frac{d}{dt}U_{\rm \mathit{rf}}\right)U_{\rm \mathit{rf}}^\dagger+U_{\rm \mathit{rf}}HU_{\rm \mathit{rf}}^\dagger\right)\ket{\phi_{\rm \mathit{rf}}}.
\end{align}

We define $H_{\rm \mathit{eff}} = \mathrm{i}(d U_{\rm \mathit{rf}}/dt)U_{\rm \mathit{rf}}^\dagger+U_{\rm \mathit{rf}}HU_{\rm \mathit{rf}}^\dagger$. $H_{\rm \mathit{eff}}$ can be viewed as the form of $H$ in the rotating frame. Substituting the specific expression of $U_{\rm \mathit{rf}}$, we obtain:
\begin{align}
    H_{\rm \mathit{eff}} 
    &= \omega^{+} \ket{\phi^+}\bra{\phi^+}\nonumber\\
    &- \frac{\Omega^-}{2}\left(\ket{00}\bra{\phi^-} + \ket{\phi^-}\bra{00}\right)\\
    &-\frac{\Omega^+}{2}\left(e^{\mathrm{i}\omega^-t}\ket{00}\bra{\phi^+} + e^{-\mathrm{i}\omega^-t}\ket{\phi^+}\bra{00}\right).\nonumber
\end{align}

Using the RWA, we can drop the third term in $H_{\rm \mathit{eff}}$, which means we can neglect the interaction between $\ket{00}$ and $\ket{\phi^+}$. Subsequently, we can also drop the first term in $H_{\rm \mathit{eff}}$ because $\ket{00}$ will not evolve into $\ket{\phi^+}$. Finally, the simplified $H_{\rm \mathit{eff}}$ can be written as
\begin{equation}
H_{\rm RWA} = -\frac{\Omega^-}{2}\left(\ket{00}\bra{\phi^-} + \ket{\phi^-}\bra{00}\right).
\end{equation}
It is clear that the excitation process for this $Q$-$C$ system when $\omega_d = \omega^-$ is just Rabi oscillations between $\ket{00}$ and $\ket{\phi^-}$.

However, the approximation is not always valid, we need to carefully assess the conditions under which the RWA will fail. For $H_{\rm \mathit{eff}}$, we move into a new rotating frame upon the previously established rotating frame. The frame rotates with $\ket{\phi^+}$ at frequency $\omega^+$. We define $H^*_{\rm \mathit{eff}}$ as the form of $H_{\rm \mathit{eff}}$ in the new rotating frame. Following the same deduction process as above, we obtain:
\begin{align}
    H_{\rm \mathit{eff}}^*  
    &= -\frac{\Omega^-}{2}\left(\ket{00}\bra{\phi^-} + \ket{\phi^-}\bra{00}\right)\nonumber\\
    &- \frac{\Omega^+}{2}\left(e^{\mathrm{i}\Delta\omega t}\ket{00}\bra{\phi^+} + 
e^{-\mathrm{i}\Delta\omega t}\ket{\phi^+}\bra{00}\right),
\end{align}
where $\Delta\omega = \omega^--\omega^+$. It is quite clear that the RWA can work well if $\Omega^+/2 \ll \Delta\omega$. Since the coupler is near resonance with the qubit, we have $\Delta\omega \approx 2g_{qc} \approx 200$ MHz. The shortest gate length of $X$ gate is 30 ns in our experiment, which means $\Omega^+/2$ is around 16 MHz. Generally, $\Omega^+/2 \ll \Delta\omega$ is well satisfied in the excitation process. So theoretically, we prove that the exciting pulse frequency and length (minimum at 30 ns) in our experiment are plausible to excite the initial state $\ket{00}$ to the dressed state $\ket{\phi^-}$.

While reading out, we can only detect the state of the qubit. Since $\ket{\phi^-}$ is a linear combination of the excited state of both the qubit and the coupler (with the qubit state dominating the population), the $\ket{1}$ probability of the qubit reaches it maximum only if $\ket{00}$ is fully excited to $\ket{\phi^-}$. If there are distortions from the Z line of the coupler, the frequency of $\ket{\phi^-}$ will change significantly due to the near resonance of the coupler with the qubit. This results in $\ket{00}$ being only partially excited to $\ket{\phi^-}$, causing a noticeable drop in the probability of qubit in $\ket{1}$. Therefore, this exciting procedure can serve as an effective criterion for characterizing the coupler's Z pulse distortion.

The distortion is measured by the compensation that maximizes the $\ket{1}$ probability of the qubit after applying the calibrated excitation pulse. It should be emphasized that the duration of the excitation pulse (30ns-200ns) is carefully selected. Firstly, the measured distortion should represent an "average" distortion during the excitation pulse period. If the duration is much longer than the relaxation time of the distortion during the exciting period, it will be incorrect to use this "average" to represent the distortion at the midpoint of the excitation pulse. Therefore, the length of the pulse should not be too long (maximum at 200 ns). Secondly, using shorter excitation pulses ensures more predictable "average" distortions over this shorter duration. However, as previously discussed, we cannot shorten the duration excessively because the RWA will fail if the duration is too short.

The feasibility of our method in calibrating the distortion is numerically validated using Python QuTiP package~\cite{Johansson2012} through the following steps:

\textbf{Step 1}: Calibrate the frequency of the excitation pulse for the qubit when the coupler Z pulse relaxes to a stable value. In our simulation, we calculate the lower eigenenergy of $H$ in Eq.~\eqref{equ1} to fit the frequency of red arrows in Fig.~\ref{fig1}(b) and (c) when setting the Z pulse amplitude (frequency) of the coupler as a constant.

\textbf{Step 2}: Calibrate the amplitude of this excitation pulse under the same condition as in Step 1. Perform the $\pi$ pulse with $H_{\rm driven} = f(t)\sigma_x$ and $f(t)$ as a Gaussian envelope. We test durations of 30ns, 100ns and 200ns, and found that the product of duration and the amplitude of maximum excitation is nearly conserved, which confirms that the carefully selected $\pi$ pulse lengths are suitable.

\begin{figure}[!htb]
\includegraphics[scale=0.19]{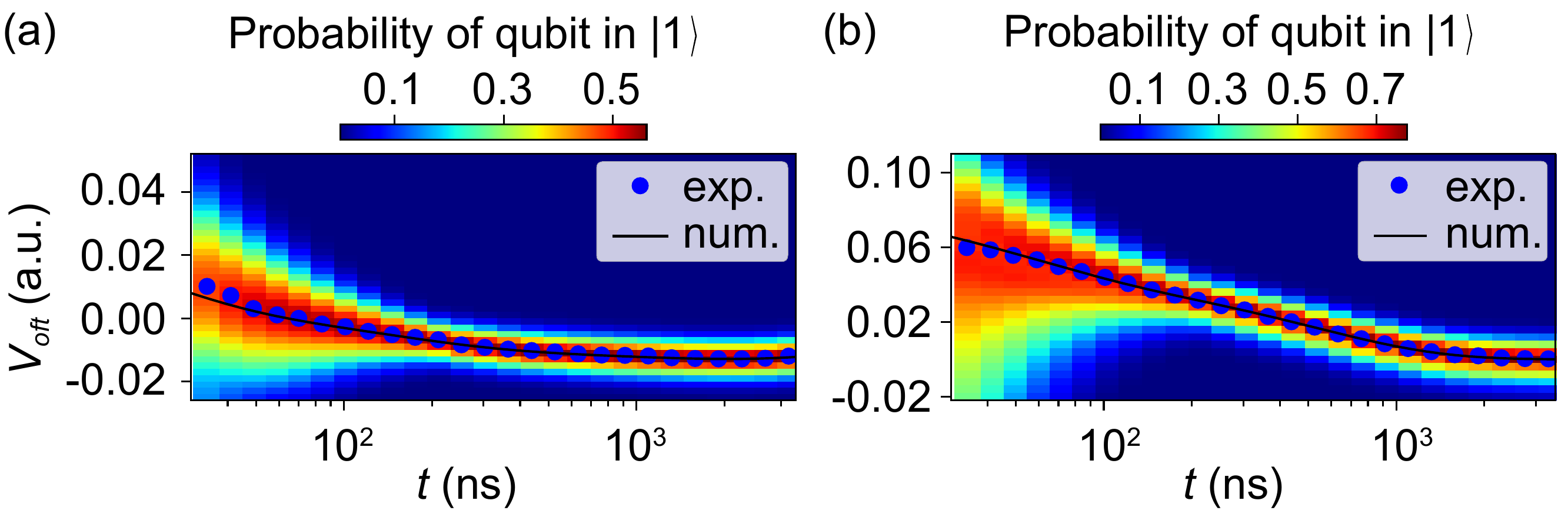}
\caption{\label{figD1}Numerical results of pulse distortion. (a) and (b) are the results for Chip1 and Chip2. The black solid curves represent ideal distortions calculated from the parameters listed in Tab.~\ref{tableC1} and Tab.~\ref{tableC2}, and the blue dots represent the results from numerical simulations.}
\end{figure}

\textbf{Step 3}: Use the response function of both chips achieved from experiments to distort the coupler's Z pulse. Calculate the frequency changes of the coupler using spectral information such as $E_c$, $E_j$, and so on. Add all the time-dependent Hamiltonians (excluding the readout pulse) shown in Fig.~\ref{fig2}(b). At a specific time $t$, detect the distortion (compared with the stable value in Step 1) by adding different Z compensation amplitudes to the coupler. Identify the specific amplitude that maximizes the probability of qubit in $\ket{1}$. We finally compare the compensation results between the ideal values calculated from the parameters listed in Tab.~\ref{tableC1} and Tab.~\ref{tableC2} and the simulated values (blue dots shown in Fig.~\ref{figD1}). The results show perfect consistency for both chips.

\section{Comparison with other experimental methods\label{appE}}
\setcounter{figure}{0}
\setcounter{table}{0}

We present three other different experimental methods to demonstrate the consistency of all the methods and the efficiency of our proposed method on Chip2. 

The predistored Z square waves below are all calculated using parameters listed in Tab.~\ref{tableC1} and Tab.~\ref{tableC2} (obtained using our method). The Z square waves are all 3000ns in length.

\begin{figure}[!htb]
\includegraphics[scale=0.19]{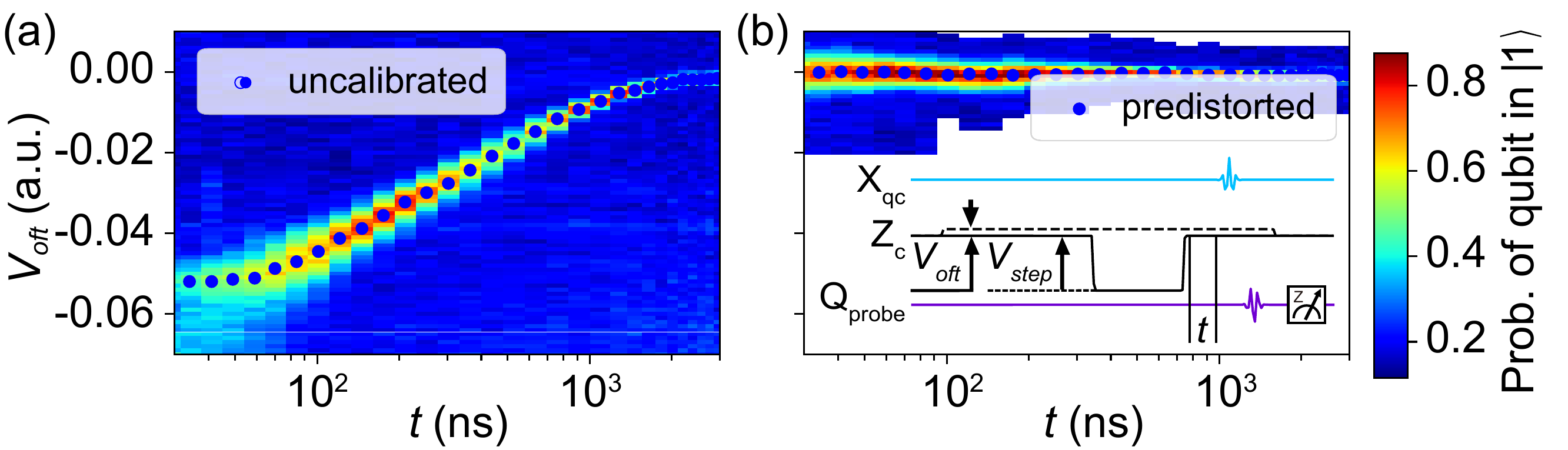}
\caption{\label{figE1}Excitation and readout of coupler via AC stark effect. (a) and (b) show the measured distortions of uncalibrated and predistored pulses, respectively. The inset of (b) illustrates the pulse sequence used for both (a) and (b). We apply a square wave with an amplitude $V_{\rm \mathit{step}}$ and add an additional offset bias $V_{\rm \mathit{oft}}$ to compensate for the distortion of this square wave. When the distortion is compensated, the crosstalk $\pi$ pulse will excite the coupler, resulting in the frequency shift of the qubit caused by the AC stark effect. Then the probe qubit will not be excited by its original $\pi$ pulse.}
\end{figure}

\textbf{Method 1}: Excitation and readout of coupler via AC stark effect~\cite{Zhang2023}. We utilize the XY line of one nearby qubit to excite the coupler by the XY-line crosstalk. The parameters of the exciting pulses are listed in Tab.~\ref{tableB2}. After the coupler is excited, the effective frequency of another nearby qubit changes due to the AC stark effect so that it can serve as a probe. Then the qubit is hard to excite with hundreds of nanoseconds pulse length, which has a smaller spectral broadening than AC stark shift. The pulse sequence for calibrating the flux distortion of the coupler after a Z square wave with amplitude $V_{\rm \mathit{step}}$ is shown in the inset of Fig.~\ref{figE1}(b). We show the probability of the qubit in state $\ket{0}$ with the change of an additional offset bias $V_{\rm \mathit{oft}}$ to compensate for the distortion of the square wave, which is different from Ref~\cite{Zhang2023}. The distortion equals the specified $V_{\rm \mathit{oft}}$ which maximizes the probability of the qubit in state $\ket{0}$. the results for uncalibrated and predistored Z square waves are shown in Fig.~\ref{figE1}, respectively. This method is disadvantaged as it requires the XY crosstalk to excite the coupler.

\begin{figure}[!htb]
\includegraphics[scale=0.19]{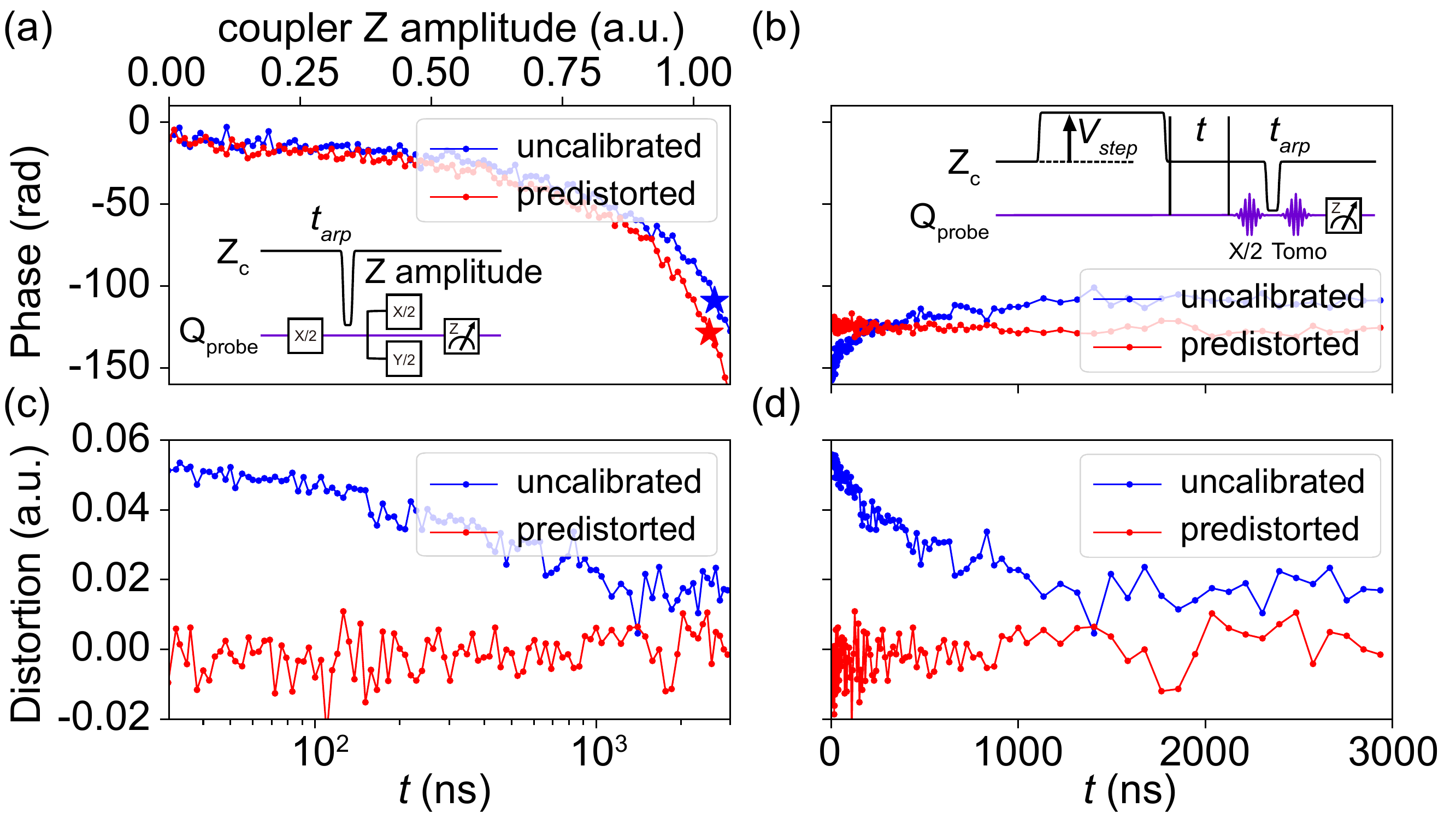}
\caption{\label{figE2}Phase detection of qubit. (a) shows the relationship between the effective coupler Z amplitude and the accumulated relative phase. The inset of (a) displays the pulse sequence used for calibrating the phase criterion. We apply a flux bias on the coupler for $t_{arp}=20$ ns to accumulate the relative phase of the probe qubit (initiated in state $(\ket{0}+\ket{1})/\sqrt{2}$). Stars represent the selected criterion and working points for (b-d). (b) shows the accumulated relative phase at time $t$ after a Z square wave with amplitude $V_{\rm \mathit{step}}$ is applied. (c) and (d) show the distortion calculated from (b) using the relationship in (a), with (c) displaying the time $t$ on a logarithmic scale. All items in blue or red represent the results for the uncalibrated or predistored Z square waves, respectively.}
\end{figure}

\textbf{Method 2}: Phase-detection of qubit~\cite{Cao2023, Zhao2024, Li2024a}. The qubit is pre-excited to state $(\ket{0}+\ket{1})/\sqrt{2}$ and we detect its relative phase using quantum state tomography (QST). The pulse sequences of the experiments are shown in the insets of Fig.~\ref{figE2}(a) and (b). First, a criterion of phase is required to be calibrated. We bias the coupler for a short duration $t_{arp}$ (20ns in our experiment), during which the qubit accumulates relative phase due to its frequency shifts caused by the change of the coupler frequency. As shown by the stars in Fig.~\ref{figE2}(a), we select a sensitive working point, where the probe qubit accumulates enough phase to change drastically with variations in effective coupler Z amplitude. Next, we apply a Z square wave with amplitude $V_{\rm \mathit{step}}$ before and then detect the qubit phase (see Fig.~\ref{figE2}(b)). For a specific time $t$, the distortion can be determined using the relationship between effective coupler Z amplitude and the accumulated relative phase from Fig.~\ref{figE2}(a). The distortion results for both uncalibrated and predistored Z square waves are shown in Fig.~\ref{figE2}(c) and (d), where (c) displays the time $t$ on a logarithmic scale. This method is disadvantaged as the relative phase is sensitive to noise, making the result unstable and noisy.

\begin{figure}[!htb]
    \includegraphics[scale=0.19]{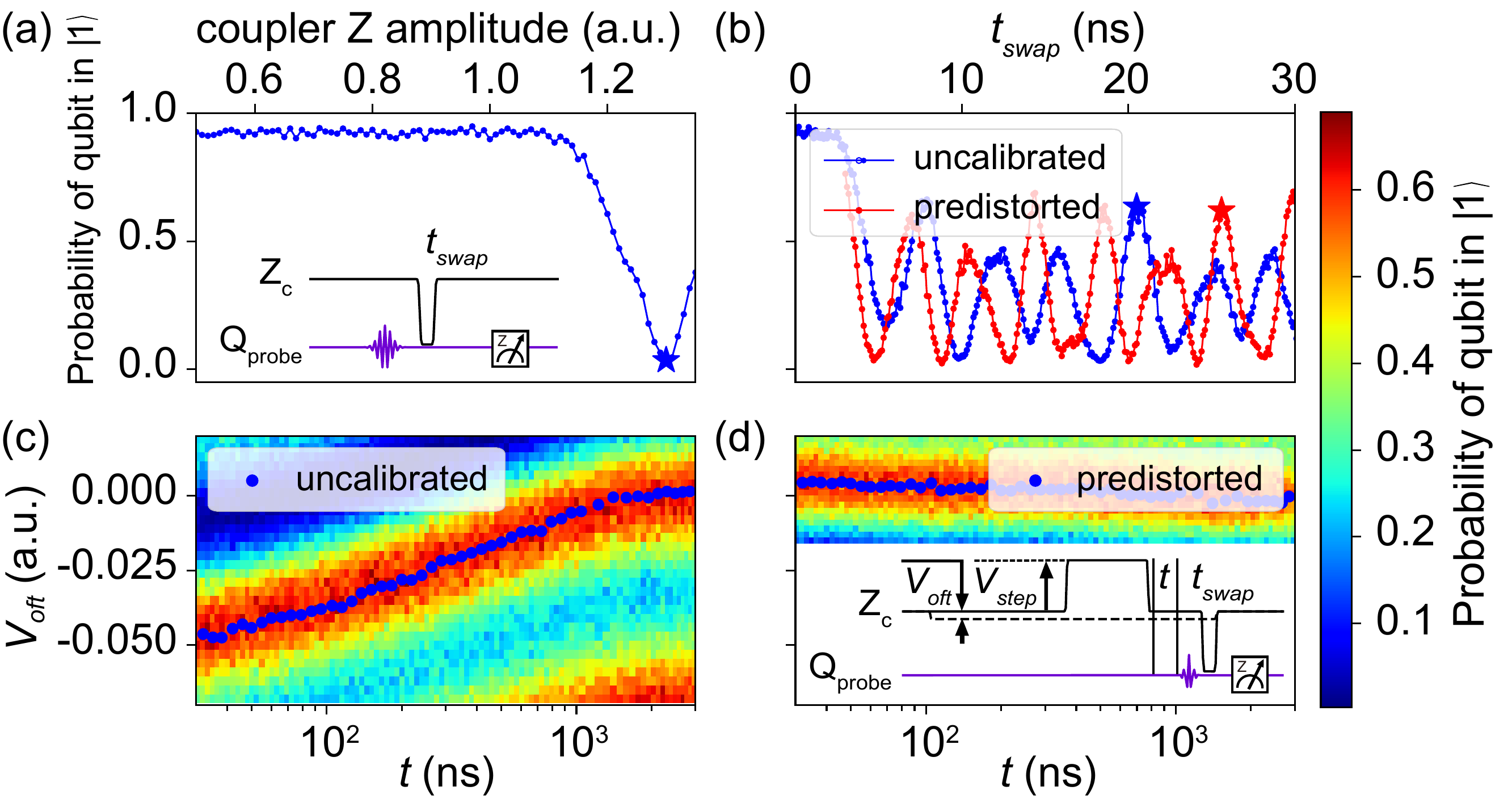}
\caption{\label{figE3}Qubit-coupler swapping. (a) The probability of qubit in $\ket{1}$ vs. the coupler Z amplitude. The inset of (a) shows the pulse sequence used for (a) and (b). We choose the blue star in (a) as the working point for the coupler in experiments (b-d). (b) The probability of qubit in $\ket{1}$ vs. the swapping time. We select a specific swapping time (the stars in (b)) as the working point for pulse calibration. (c) and (d) show the distortion for the uncalibrated and predistored Z square wave, respectively. The inset of (d) shows the experimental sequence used for (c) and (d). Z square wave with amplitude $V_{\rm \mathit{step}}$ is applied before the calibrated swapping pulses, and an additional offset bias $V_{\rm \mathit{oft}}$ compensates for the distortion of this square wave.}
\end{figure}

\textbf{Method 3}: Qubit-coupler swapping. The qubit is excited, and the coupler is biased to resonate and swap population with the qubit. First, we fix the swapping time and then adjust the coupler Z amplitude to find the resonating working point for the coupler (sequence shown in the inset of Fig.~\ref{figE3}(a)). The population of the qubit mainly swaps into the coupler at the star point in Fig.~\ref{figE3}(a). Next, at the calibrated working point, we select a specific swapping time (swapping process shown in Fig.~\ref{figE3}) when the population of the qubit is at its peak, severing as the working point for pulse calibration (stars shown in Fig.~\ref{figE3}(b)). Then, a Z square wave with amplitude $V_{\rm \mathit{step}}$ is applied before the calibrated swapping process, and an additional offset bias $V_{\rm \mathit{oft}}$ is added to compensate for the distortion of this square wave (sequence shown in the inset of Fig.~\ref{figE3}(d)). The population of the qubit is maximized when the output after compensating with $V_{\rm \mathit{oft}}$ meets the criterion we calibrated before. This specific $V_{\rm \mathit{oft}}$ represents the distortion of the square wave at the specific time $t$. The distortion for both uncalibrated and predistored Z square waves are shown in Fig.~\ref{figE3}(c) and (d), respectively. This method is disadvantaged due to its poor resolution and multi-peak results (shown in data especially when $t>$1000ns in Fig.~\ref{figE3}(c)).

\section{Two-qubit gates and cross-entropy benchmarking\label{appF}}
\setcounter{figure}{0}
\setcounter{table}{0}

\begin{figure*}[!htb]
    \includegraphics[scale=0.35]{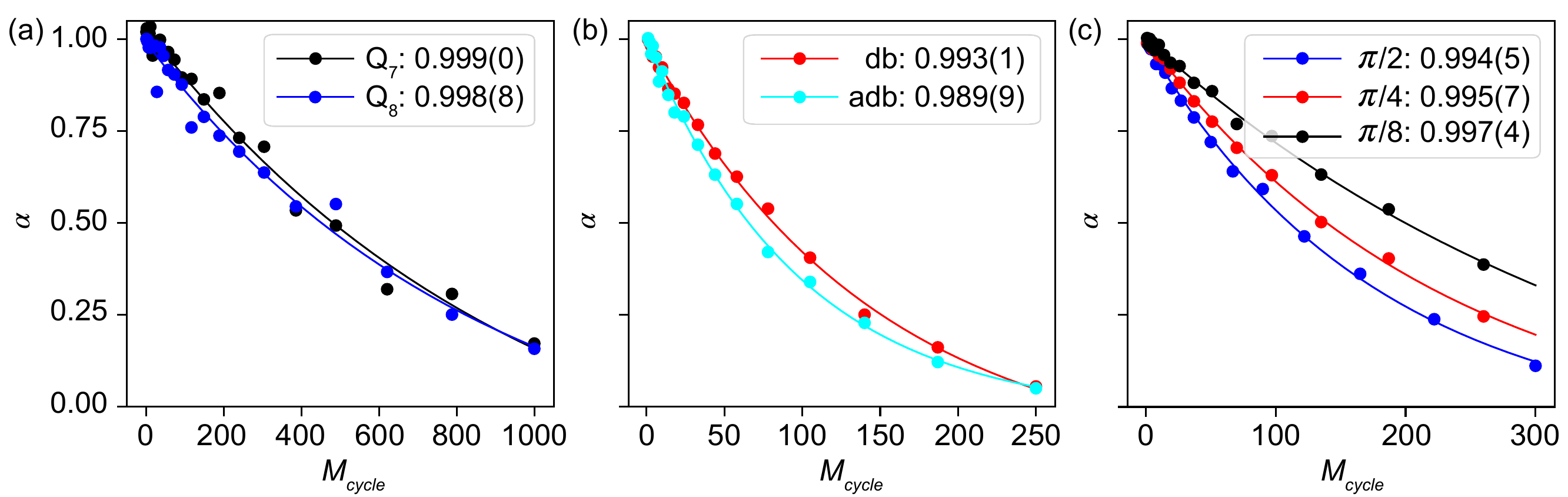}
\caption{\label{figF1}Fidelities results from XEB benchmarking for (a) single-qubit, (b) CZ ("db" for diabatic, "adb" for adiabatic), and (c) CPhase gates. Fidelities $\alpha$ are obtained from $k=40$ random sequences.}
\end{figure*}

The durations of the adiabatic and diabatic CZ gates are 45 and 36 ns, respectively. The fidelities of both CZ gates, as discussed in Section~\ref{sec4}, are also measured using cross-entropy benchmarking (XEB)~\cite{Boixo2018, Arute2019}. The two single-qubit benchmarks are performed simultaneously, and results are illustrated in Fig.~\ref{figF1}(a). The CZ gate is interleaved within the cycles of the simultaneous random single-qubit gates. The XEB benchmarking results are shown in Fig.~\ref{figF1}(b). The average XEB fidelities of the adiabatic and diabatic CZ gates are $F_{\rm XEB}=99.37\pm0.04\%$ and $F_{\rm XEB}=99.61\pm0.04\%$, respectively, which are consistent with the average RB fidelities.

Under the same single-qubit conditions, we also perform a set of CPhase gates with conditional phases of $\pi/2$, $\pi/4$ and $\pi/8$ and with effective lengths of 32, 32 and 27 ns, respectively. These CPhase gates are implemented using a waveform similar to the diabatic CZ gate in Fig.~\ref{fig3}(a). The key difference lies in the larger energy gap between states $\ket{11}$ and $\ket{20}$ during the coherent exchange, which ensures reduced geometry phase accumulation compared with the resonance condition while maintaining an acceptable level of leakage. The average XEB fidelities of these CPhase, as shown in Fig.~\ref{figF1}(c), are $99.71\pm0.03\%$, $99.81\pm0.03\%$ and $99.94\pm0.04\%$, respectively.

The fidelities $\alpha$ of repeated gate cycles are calculated using the cross entropy between the experimental probability distribution obtained from $k=40$ random circuits and the ideal output probability of these circuits. The errors of cycle fidelities shown in the legends represent the fitting errors. 

The duration of iSWAP gate is 24 ns. The average XEB fidelity $F_{\rm XEB}=99.82\pm0.02\%$ of this iSWAP gate is calibrated using $k=50$ random circuits. The fitting uncertainty of all $F_{\rm XEB}$ (CZ, CPhase, iSWAP gates) are calculated using the equations outlined in Appendix~\ref{appG}.

\section{Derivations of fitting errors for RB and XEB benchmarking\label{appG}}

All the fidelities discussed in the main text are errors included. The uncertainty in the legends of all figures indicates the fitting errors of the fidelity. The fitting function used is $F(n)=Ap^n+B$, where $F$ represents the sequence fidelities after $n$ gate cycles, $1-p$ is the rate of depolarization, and the parameters $A$ and $B$ capture the state preparation and measurement errors~\cite{Chen2018}. The errors in depolarization fidelities (see the legends in Fig.~\ref{fig3}(d)) are fitting errors~\cite{Xu2021, Li2024a}, calculated as "perr" of $p$ gaining from the return value "pcov" of Python scipy.optimize.curve\_fit function with the relation perr=sqrt(diag(pcov)) and defined as $\sigma_p$. Furthermore, the uncertainties in gate fidelities $\sigma_F$ are calculated using the law of propagation of uncertainties.

The RB gate fidelity~\cite{Chen2018} is related to the depolarization fidelities $p_{\rm ref}$ for reference Clifford sequence and $p_{\rm gate}$ for interleaved CZ gate sequence, as follows:
\begin{equation}
    F_{\rm RB} = 1-\left(1-\frac{p_{\rm gate}}{p_{\rm ref}}\right)\frac{D-1}{D},
\end{equation}
where $D=2^n=4$ is the dimensionality of the Hilbert space. The error $\gamma = 1-F_{\rm RB}$ represents the average error. The uncertainty of $F_{\rm RB}$ is calculated as
\begin{align}
    \sigma_{F_{\rm RB}} 
    &= \sqrt{\left(\frac{\partial F_{\rm RB}}{\partial p_{\rm gate}}\right)^2\sigma_{p_{\rm gate}}^2 + \left(\frac{\partial F_{\rm RB}}{\partial p_{\rm ref}}\right)^2\sigma_{p_{\rm ref}}^2}\nonumber\\
    &= \frac{3}{4}\frac{p_{\rm gate}}{p_{\rm ref}}\sqrt{\left(\frac{\sigma_{p_{\rm gate}}}{p_{\rm gate}}\right)^2 + \left(\frac{\sigma_{p_{\rm ref}}}{p_{\rm ref}}\right)^2}.
\end{align}

The XEB gate fidelity~\cite{Arute2019} is related to the depolarization fidelities $p_{Q_1}$ and $p_{Q_2}$ for two single-qubit XEB sequences and $p_{\rm gate}$ for interleaved CZ XEB sequence. A typical gate cycle contains two parallel single-qubit gates and a serial CZ gate if interleaved. The fidelity for parallel gates is calculated as the product of $F_{Q_i} = 1-\gamma_{Q_i} = 1-(1-p_{Q_i})(1-1/D_1^2)$ ($i=1,2$) of both single-qubit gates, where $\gamma_{Q_i}$ is the Puali error and $D_1=2$. Then the depolarization rate for parallel single-qubit gates is $1-p_{\rm sq} = (1-F_{Q_1}F_{Q_2})/(1-1/D_2^2)$, where $D_2=4$. After simplifying, we have
\begin{equation}
    p_{\rm sq} = \frac{p_{Q_1}+p_{Q_2}+3p_{Q_1}p_{Q_2}}{5}.
\end{equation}
The uncertainty of $p_{\rm sq}$ is calculated as
\begin{align}
    \sigma_{p_{\rm sq}} 
    &= \sqrt{\left(\frac{\partial p_{\rm sq}}{\partial p_{Q_1}}\right)^2\sigma_{p_{Q_1}}^2 + \left(\frac{\partial p_{\rm sq}}{\partial p_{Q_2}}\right)^2\sigma_{p_{Q_2}}^2}\nonumber\\
    &= \sqrt{\left(\frac{A_{Q_2}}{5}\right)^2\sigma_{p_{Q_1}}^2 + \left(\frac{A_{Q_1}}{5}\right)^2\sigma_{p_{Q_2}}^2},
\end{align}
where define $A_{Q_i}=1+3p_{Q_i}$ ($i=1,2$).
The fidelity for serial gates is calculated as the product of depolarization fidelities, so the average XEB gate fidelity~\cite{Li2022a} is defined as:
\begin{equation}
    F_{\rm XEB} = 1-\left(1-\frac{p_{\rm gate}}{p_{\rm sq}}\right)\frac{D-1}{D},
\end{equation}
where the error $\gamma = 1-p_{\rm gate}/p_{\rm sq}$ represents the depolarization error, and coefficient $(D-1)/D$ comes from the transformation from depolarization error to the average error~\cite{Arute2019, Barends2019a}. The uncertainty of $F_{\rm XEB}$ is calculated as
\begin{equation}
    \sigma_{F_{\rm XEB}} = \frac{3}{4}\frac{p_{\rm gate}}{p_{\rm sq}}\sqrt{\left(\frac{\sigma_{p_{\rm gate}}}{p_{\rm gate}}\right)^2 + \left(\frac{\sigma_{p_{\rm sq}}}{p_{\rm sq}}\right)^2}.
\end{equation}

\end{document}